\documentclass[
  journal=largetwo,
  manuscript=article-type,
  year=2020,
  volume=37,
]{cup-journal}
\usepackage{rotating}
\usepackage{tablefootnote}
\usepackage{amsmath}
\usepackage[nopatch]{microtype}
\usepackage{hyperref}
\usepackage{url}
\usepackage{booktabs}
\title{AstroSat View of Transient Low-mass X-ray Binary XTE J1701-462: Spectral and Temporal
Evolution along the Z-track}

\author{Vivek K. Agrawal}
\affiliation{Space Astronomy Group, U. R. Rao Satellite Centre, Bangalore, 560037, Karnataka, India}
\email[Vivek K. Agrawal]{vivekag@ursc.gov.in}


\begin{document}

\begin{abstract}
\textit{AstroSat} observed  transient neutron star low-mass X-ray binary XTE J1701-462 for a total duration of $\sim$ 135 ks during its 2022 outburst. We report the results of a detailed spectral and timing analysis carried out using this data. The source traced a complete `Z' shaped structure in the hardness intensity diagram (HID). The source exhibited an extended horizontal branch and a short-dipping flaring branch in the HID. The spectra of the source were fitted with different approaches. We find that most suitable spectral model comprises emission from a standard multi-color accretion disk (diskbb in XSPEC) and Comptonized radiation from a hot central corona,  described by \textit{Comptb} model of XSPEC. The observed disk component is cool, having a temperature in the range of $\sim 0.28-0.42$ keV and truncated far ($\sim$ 250 - 1600 km)  from the compact object. The Compton corona has an optical depth in the range of $\sim 3.4- 5.1 $ and a temperature in the range of $3.3-4.5$ keV.  The disk and corona flux as well as truncation radius vary significantly along the HID. The temperature $kT_{in}$ depends on both luminosity and inner disk radius and hence shows marginal variation as compared to the truncation radius.  We discuss possible scenarios to explain the relationship between the spectral evolution and motion of the source along the HID.  The timing analysis revealed horizontal branch oscillations (HBOs) in the frequency range $\sim 34-40$ Hz. The frequency and rms strength of HBO vary systematically as the source moves along the horizontal branch (HB). The observed correlation of the HBO properties with the position on the HB is similar to that previously reported in this source using \textit{RXTE} data during the 2006 outburst of the source. The source also showed  normal branch oscillations (NBOs) with frequency $\sim$ 6.7 Hz in the middle and the lower normal branch. The energy-dependent study of the HBO properties suggests that the HBO is stronger in the higher energy band. We also observed very-low frequency noise (VLFN) and band-limited noise (BLN) components in the power density spectra. The break frequency of BLN component was found to be tightly correlated with the HBO frequency. We discuss possible models to explain the origin and nature of the observed features in the PDS.
\end{abstract}

\section{Introduction}
A low magnetic field neutron star (NS) accreting material from a low mass companion star  provides a way to understand the complex accretion and emission processes close to the ultra-dense star. X-ray flux and spectral-timing properties of neutron star low-mass X-ray binaries (NS-LMXBs) vary on time scales of hours to months causing  them to trace different patterns on the color-color (CCD) and hardness-intensity diagram (HID). The most luminous and persistent NS-LMXBs, known as Z-sources trace a pattern that resembles an approximate `Z'  shape.  Three distinct branches  of the Z-pattern are  horizontal branch (HB), normal branch (NB) and flaring branch (FB). Major fractions of the NS-LMXBs trace a  fragmented  pattern,  consisting of a curved banana branch and an island type structure \citep{1989A&A...225...79H}. Z-sources have luminosity  in the range of 0.5$-$1.0 L$_{Edd}$, while luminosity of atoll sources vary in  the range of 0.01-0.2 L$_{Edd}$ \citep{2001AdSpR..28..307B}. It has been argued that  mass accretion rate increases as Z-sources move from the HB to the NB and then to the FB \citep{1990A&A...235..131H}. There are six Galactic persistent Z-sources \citep{1989A&A...225...79H}, one extra-galactic Z-source \citep{2000ApJ...528..702S}and one transient Z-source \citep{2007ApJ...656..420H}. GX 13+1 and Cir X-1  is also considered as candidate Z-sources \citep{1999ApJ...517..472S,2003A&A...406..221S, 2015ApJ...809...52F}. The neutron star transient source IGR J17480-2446 also showed both `Z' and `atoll' like behavior \citep{2011MNRAS.418..490C}.

The energy spectra of Z-sources are soft in nature  and consist of  two main emission components.  Most prominent components are, a cool Comptonized emission from a corona and a thermal emission from a multi temperature accretion disk or a boundary-layer around the NS \citep{2001ApJ...554...49D, 2002A&A...386..535D,2004NuPhS.132..608I, 2020Ap&SS.365...41A,2020MNRAS.497.3726A}. An exact nature of the soft component, geometry  and location of the corona that produces the Comptonized emission by inverse Compton process are still under debate. In some cases, other than two main spectral components, a signature of reflection of the coronal emission from an  accretion disk \citep{2018ApJ...867...64C,2022ApJ...927..112L}  is also seen.  A high energy tail is also found to be present in the spectra of the Z-sources \citep{2001ApJ...554...49D,2002A&A...386..535D,2020Ap&SS.365...41A}. The atoll sources exhibit two spectral states, high-soft and low-hard. In the high-soft state the spectra of these sources resemble those of Z-sources. However in the low-hard state the spectra has Comptonized component with temperature $\sim$10$-$30 keV \citep{2000ApJ...533..329B,2002MNRAS.337.1373G}.

Low Frequency Quasi-periodic Oscillations (LFQPOs) in the frequency range $5-70$ Hz have been observed in the power density spectra (PDS) of Z-sources \citep{2006csxs.book...39V}.  In Z-sources, the centroid frequency and other properties of the LFQPOs show a strong dependence on  the  position  of the source on the HID and CCD.  Horizontal branch oscillations (HBOs) with a frequency $\sim 10-70$ Hz, normal branch oscillations (NBOs) with a frequency $5-10$ Hz and flaring branch oscillations (FBOs) with centroid frequency $\sim$ 20 Hz are seen in the HB, NB  and FB respectively \citep{2002ApJ...568..878H,2002MNRAS.333..665J}. FBOs and NBOs are connected and possibly have a same origin. Many competing models exists to explain the origin of the LFQPOs. However, the mechanism that can explain all the observed properties of these features is not fully understood.

XTE J1701-462  is the first Z-source that shows transient behavior \citep{2007ApJ...656..420H}. It has undergone two outbursts, the first one on 2006 January 18 \citep{2006ATel..696....1R}. The second outburst was detected on 2022 September 6 by MAXI/GSC \citep{2022ATel15592....1I}. During the first outburst the source evolved from a Cyg-like Z-source to a Sco-like Z-source and then finally at the end of the outburst it displayed an atoll-like behavior \citep{2009ApJ...696.1257L}.  Main difference between the Cyg-like and the Sco-like  Z-sources are the shape of their Z-tracks. The Cyg-like sources have an extended  and a horizontal HB, while the  Sco-like  Z-sources have a slanted HB. \citet{2009ApJ...696.1257L} used a model consisting of a single temperature black-body (BB), a multi-color-disk-blackbody (MCD) emission and a constrained broken power-law (CBPL) to describe the X-ray spectra of this source during various stages of its outburst. A detailed spectral evolution study was performed  by \citet{2011AJ....142...34D} employing the model used by \citet{2009ApJ...696.1257L}. Three type-I X-ray bursts were reported in the source during 2006-2007 outburst \citep{2009ApJ...699...60L}. Using the  photospheric expansion burst a distance to the source was estimated to be 8.8 kpc \citep{2009ApJ...699...60L}. The HBOs with a frequency $\sim$ 10$-$60 Hz and the NBOs with a frequency $\sim$ 7-10 Hz have been observed in the source \citep{2010ApJ...719..201H,2007ApJ...656..420H}. Moreover, pairs of kHz QPOs  are observed during the Z-stage of the source \citep{2010MNRAS.408..622S}. X-ray polarization with a polarization degree $\sim$ 4.5 \% and a polarization angle $\sim$ 143$^\circ$ has been detected in the source. \citep{2023A&A...674L..10C,2023MNRAS.525.4657J}.  The recent X-ray spectro-polarimetric study carried out using the simultaneous data from \textit{IXPE}, \textit{NuSTAR} and \textit{Insight-HXMT} showed that degree of polarization decreases as the source moves from HB to NB \citep{2024arXiv240102658Y}.
 
In this work, we have carried out spectral and temporal studies of the low-mass X-ray binary XTE J1701-462 during its 2022 outburst using the \textit{AstroSat} data. The HID revealed a pronounced HB and a dipping FB. A detailed spectral and temporal evolution study has been carried out to understand the origin of the Z-pattern and the power-spectral features. The HBOs and NBOs are found to be present in the power density spectra. The remainder of the paper is arranged in the following manner. Section \ref{sec:obs} gives the details of observation and the procedures of data reduction.  Section \ref{sec:dna} provides a description of the methods adopted  to carry out spectral and temporal data analysis. Results of the analysis are presented in the section \ref{sec:result}. Finally, we explain the  results obtained by the analysis in section \ref{sec:discus}.

\section{Observation and Data Reduction} \label{sec:obs}
\begin{figure*}[]
\centering\includegraphics[width=0.8\textwidth]{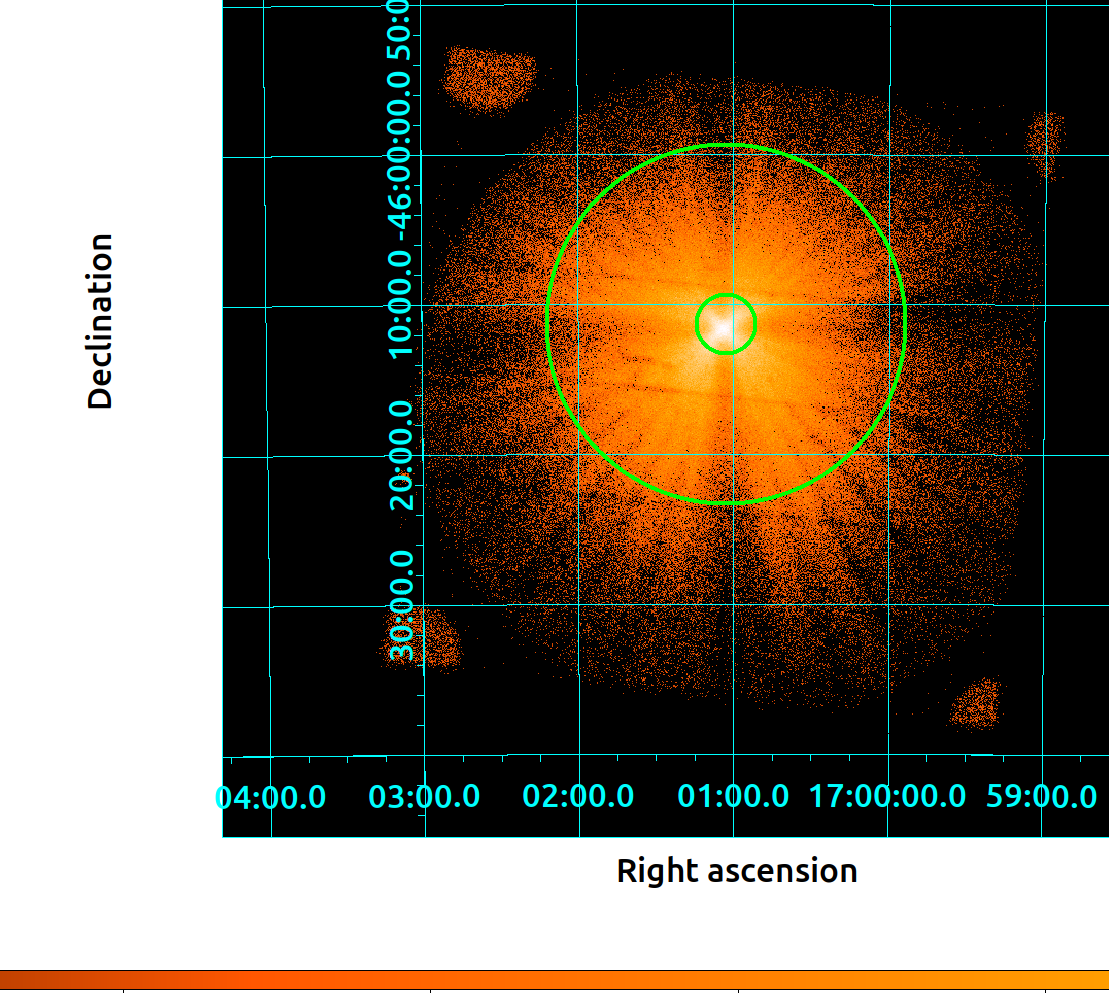}
\caption{The figure shows the \textit{SXT} image of the source in the $0.3-8$ keV energy band. The bright spots at the four corners are image of Fe$^{55}$ calibration sources. The annular region used to extract the lightcurves and spectra is also shown. For details see the text. \label{fig:sxt-image}}
\end{figure*}

{\textit AstroSat} observed the source XTE J1701-462 from 2022 September 25   to 2022 September 26 (Obs A)  and from 2022 October 1 to 2022 October 3  (Obs B)  during its 2022 outburst. The source was observed for a net exposure time of 50 ks during Obs A and 85 ks during Obs B.  The data collected from Soft X-ray Telescope (\textit{SXT}) and Large Area X-ray Proportional Counters (\textit{LAXPC})  on-board \textit{ AstroSat} were used in the present work. {\textit SXT}  carries X-ray optics and a focal plane imager that is sensitive in the energy band $0.3-8.0$ keV \citep{2016SPIE.9905E..1ES}. There are three {\textit LAXPC} units: {\textit LAXPC10}, {\textit LAXPC20} and {\textit LAXPC30}. These units operate in the energy range $3-80$ keV \citep{2016SPIE.9905E..1DY}. During the observation, {\textit LAXPC} units were operated in the event analysis mode that has time tagging accuracy of 10 micro seconds. \textit{SXT} has two different operational modes, photon counting mode and fast-window (FW) mode. In the photon counting mode the entire frame is read out in 2.38s whereas in the FW mode the central 150 x 150 pixels are selected and read out in 278 ms. During Obs A, {\textit SXT} was operated in the photon counting mode and the FW mode was used to collect data during the  Obs B.
{\textit LAXPC} data reduction  was performed using the latest version of software package {\tt\string LAXPCSOFT}\footnote[1]{\href {https://www.tifr.res.in/~astrosat_laxpc/LaxpcSoft.html}{https://www.tifr.res.in/~asrosat\_laxpc/LaxpcSoft.html}}. This software creates standard products such as lightcurves and spectra using {\textit LAXPC} level-1 products. {\tt\string XSELECT} software version 2.5 is used to generate the image, spectra and lightcurves from {\textit SXT} Level-2 data. The count rate of the source exceeds the pileup limit ($>$ 40 counts/s; see \cite{2021MNRAS.506.6203M}) and hence we exclude events from the central 2 arc minute of the \textit{SXT} image.  We extract the \textit{SXT} spectra  from an annular region with inner radius of 2 arc-minute and outer radius of 12 arc-minute. The \textit{SXT} image marked  with selection region has been shown in  Figure \ref{fig:sxt-image}. The latest version of the response matrices are used to analyze the spectra of {\textit LAXPC} and {\textit SXT}. The Ancillary Response File (ARF) for the {\textit SXT} is created using the task {\tt\string sxtARFModule}\footnote[2]{ \href{https://www.tifr.res.in/~astrosat_sxt/dataanalysis.html}{https://www.tifr.res.in/~astrosat\_sxt/dataanalysis.html}}.  

\section{Data Analysis}\label{sec:dna}
\subsection{Generation of HID and Lightcurve} \label{sec:ccds}
\begin{figure*}[]
\centering\includegraphics[width=0.7\textwidth]{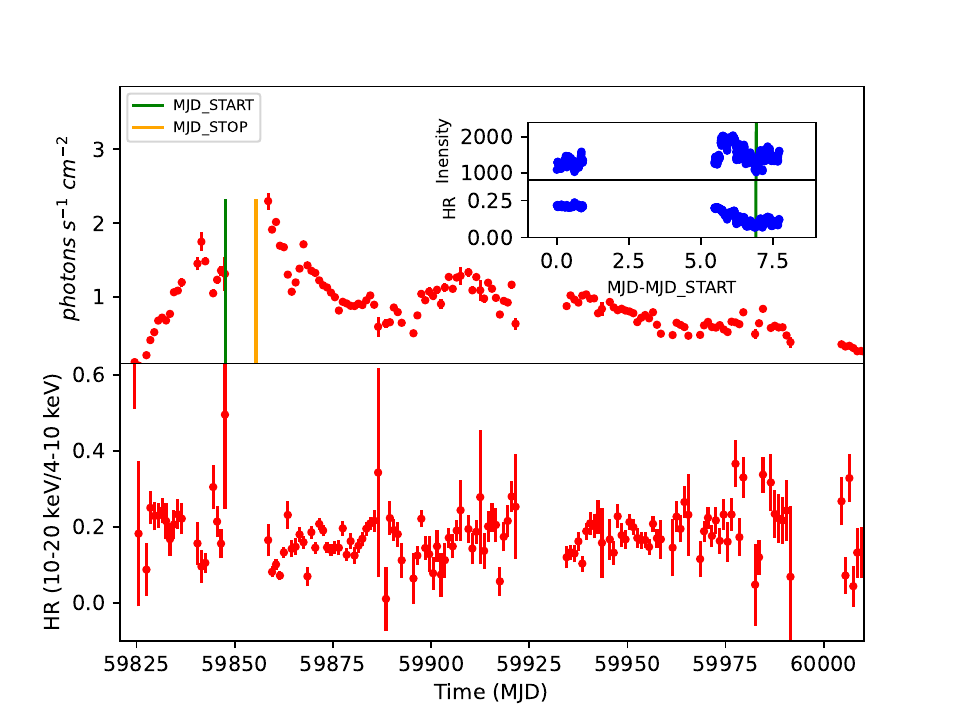}
\caption{The figure shows the \textit{MAXI} lightcurve of the source in the energy band $2-20$ keV and hardness ratio calculated using energy band 10-20 keV and 4-10 keV. The two vertical line marks the \textit{AstroSat} observations. The inset figure in top panel shows the LAXPC lightcurve in $3-60$ keV energy band and hardness ratio calculated using energy bands 9.7-20 keV and 4.6-9.7 keV. In the inset figure intensity is defined as counts rate in the 3-60 keV energy band. The start and stop time  of  the dipping FB is marked by two vertical lines in the inset figure. Since the dip period is short ($\sim$ 3000 seconds)  both vertical lines are merged.} \label{fig:maxi} 
\end{figure*}

\begin{figure*}[]
\centering\includegraphics[width=0.8\textwidth]{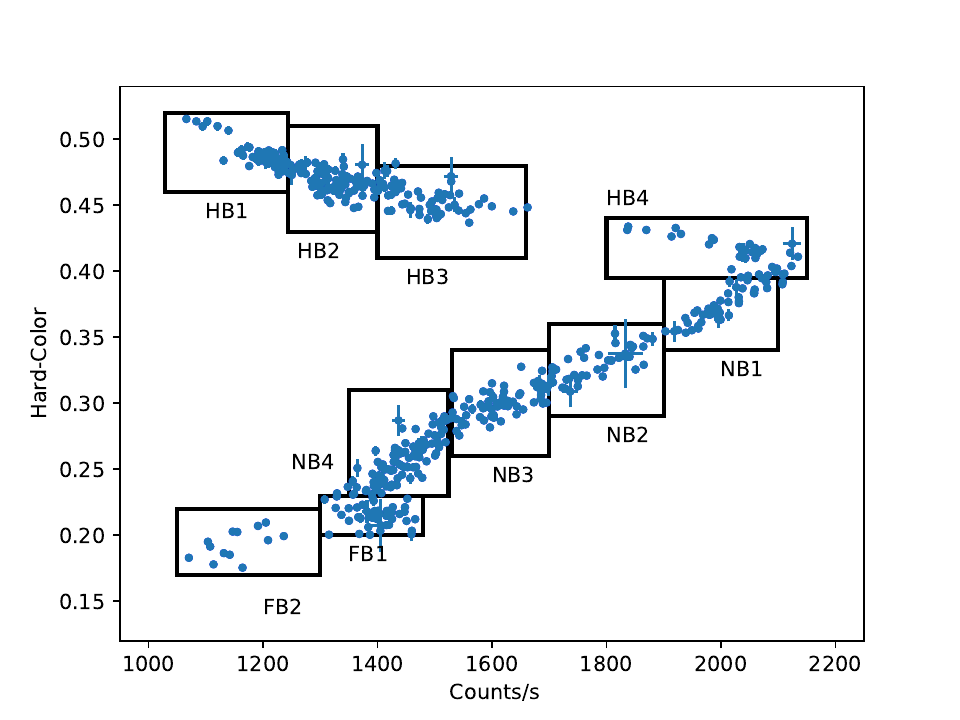}
\caption{The figure shows the HID of the source. Hard-color is ratio of the count rates in the energy bands $9.7-20$ keV and $6.5-9.7$ keV and intensity is defined as count rate in the energy range $3-20$ keV. \label{fig:hid}} 
\end{figure*}

\begin{figure*}[]
\centering\includegraphics[width=0.8\textwidth]{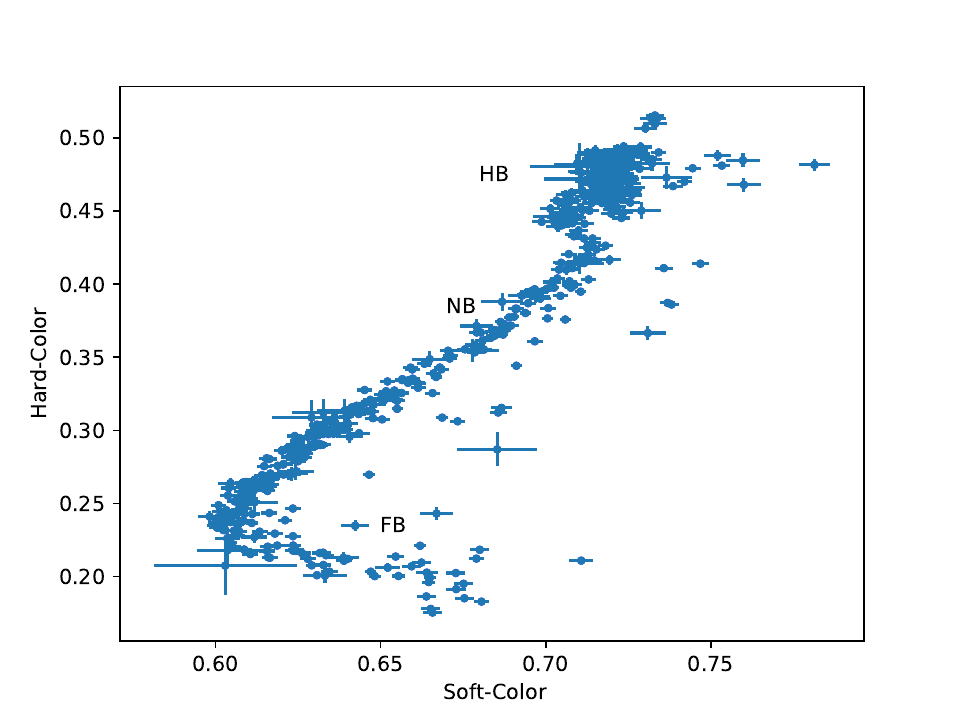}
\caption{The figure shows the CCD of the source. Hard-color is ratio of the count rates in the energy bands $9.7-20$ keV and $6.5-9.7$ keV and soft color is defined as count rate in the energy range $4.6-6.5$ and $3.0-4.6$ keV. \label{fig:ccd}} 
\end{figure*}

Since the \textit{LAXPC20} unit has  a stable gain, we used the events from this unit to create the lightcurve and HID. The task {\tt\string laxpcl1} of {\tt\string LAXPCSOFT} package is used to generate the source and background lightcurves as well as spectra. The background level for  LAXPC20 in the energy band $3-30$ keV during our observations was $\sim$ 70-80 counts/s. The time bin size used to extract the lightcurve is 256s. We have combined both Obs A and Obs B while generating the lightcurve. We  have plotted the  \textit{MAXI } lightcurve in the energy band $2-20$ keV in Figure \ref{fig:maxi}. In Figure \ref{fig:maxi}, the LAXPC lightcurve  in the energy band $3-60$ keV and variation in the LAXPC hardness ratio  are  also shown as inset. Here, the LAXPC hardness ratio is defined as ratio of count rates in the energy bands $9.7-20$ keV and 4.6-9.7 keV.  The Obs A and Obs B are marked with two vertical lines in the figure. In addition we have also plotted the hardness ratio for the \textit{MAXI} observations and LAXPC observations. It is clear from the Figure \ref{fig:maxi}  that both \textit{AstroSat} observations took place close to peak of the outburst. We have also marked the position of dipping FB (FB2) in the inset figure using vertical lines. The hardness ratio  for the \textit{MAXI} observations are defined as ratio of count rates in the energy bands $10-20$ keV and $4-10$ keV.  The \textit{LAXPC20} lightcurves in the energy ranges $6.5-9.7$ keV, $9.7-20$ keV and $3-20$ keV are used to  generate the HID. In the HID ratio of the count rates in the  energy bands $9.7-20$ keV and $6.5-9.7$ keV are plotted against intensity in the $3-20$  keV energy band. In Figure \ref{fig:hid} we show the HID of the source, where each point corresponds to a 256s bin size. HID has an extended HB and a less prominent dipping FB.  The different regions of the Z-track are marked with boxes. We extract the  spectra and the power-density spectra (PDS) corresponding to these sections to study their evolution along the HID.  The HB has been divided in four segments: HB1,HB2, HB3, HB4. The NB has four sections: NB1, NB2, NB3, NB4. The  FB is divided in FB1 and FB2. We also created CCD using lightcurves in energy bands $3.0-4.6$ keV, $4.6-6.5$ keV, $6.5-9.7$ keV and $9.7-20$ keV. The soft colors are calculated by taking ratio of the count rate in the energy bands $4.6-6.5$ keV and $3.0-4.6$ keV and the hard color is computed by taking ratio of the count rate in the energy bands $6.5-9.7$ keV and $9.7-20$ keV bands. We show the CCD of the source in Figure \ref{fig:ccd}.\\

\subsection{Analysis of Spectroscopic data}\label{sec:specana}
X-ray spectra using the \textit{LAXPC20} data and the  \textit{SXT}  data were extracted for all the ten segments of the HID. These spectra were modeled and analyzed using {\tt \string XSPEC} version 12.13.  We consider the \textit{LAXPC20} data up to 30 keV for the  spectral fitting due to limited statistics above 30 keV and dominance of  the background at higher energies. The \textit{SXT}  data in the energy range $0.7-7$ keV are used for the  combined spectral fitting. The combined broad band spectra in the energy band $0.7-30$ keV are fitted with the three different approaches.  First we used the combination of a MCD (\textit{diskbb}  in  {\tt\string XSPEC}) and a simple blackbody (BB) emission (\textit{bbodyrad} in {\tt\string XSPEC}). The model MCD+BB (hereafter \textbf{Model 1}) has been used previously to model the spectra of this source  
 \citep{2009ApJ...696.1257L} and emission in the soft state of two atoll sources \citep{2007ApJ...667.1073L}. 
 We also added a constrained broken power-law (CBPL)  component to the Model 1 following \citet{2009ApJ...696.1257L}. The photon index ($\Gamma_1$) was frozen to a value 2.5 and break energy ($E_{break}$) was fixed at 20 keV. The second power-law index was frozen to the best fit values. Only normalization of CBPL was left free. The model improves the fit for HB1, HB2 and HB3. However, for other sections of the Z-track addition of this component did not improve the fit. Hence  CBPL was not added in Model 1 for these sections.
 
 We also tried  BB plus  a Comptonized emission (hereafter \textbf{Model 2}). We used the \textit{nthComp} component in  {\tt\string XSPEC} (see  \citealt{1996MNRAS.283..193Z} for details) to describe the Comptonized emission. Finally, we attempted a combination of MCD  and  \textit{nthComp} (hereafter \textbf{Model 3}) components  to describe the X-ray spectra of this source. Both models have been widely used to fit the spectra of the NS-LMXBs \citep{2001ApJ...554...49D, 2001AdSpR..28..307B, 2002A&A...386..535D,2003MNRAS.346..933A,2023MNRAS.518..194A}. We also need an additional  smeared edge component (at $\sim$ $7-8$ keV)  to improve the fit. \textbf{Model 2} does not require this additional smeared edge component. \textit{TBabs} model \citep{2000ApJ...542..914W} is used to take into account the absorption of  X-rays  in the inter stellar medium.  We added 2 per cent systematic error while carrying out the joint spectral fitting of the \textit{LAXPC20} and \textit{SXT} spectra in {\tt \string XSPEC} to account for uncertainty   in the response matrices of both detectors. We also added a multiplicative  constant component  to account for the cross calibration related uncertainties. 
We take shape of the seed photon spectrum  as blackbody while fitting the data with \textbf{ Model 3} and \textit{diskbb} for \textbf{Model 2}.

\begin{figure*}[]
\includegraphics[width=1.0\textwidth]{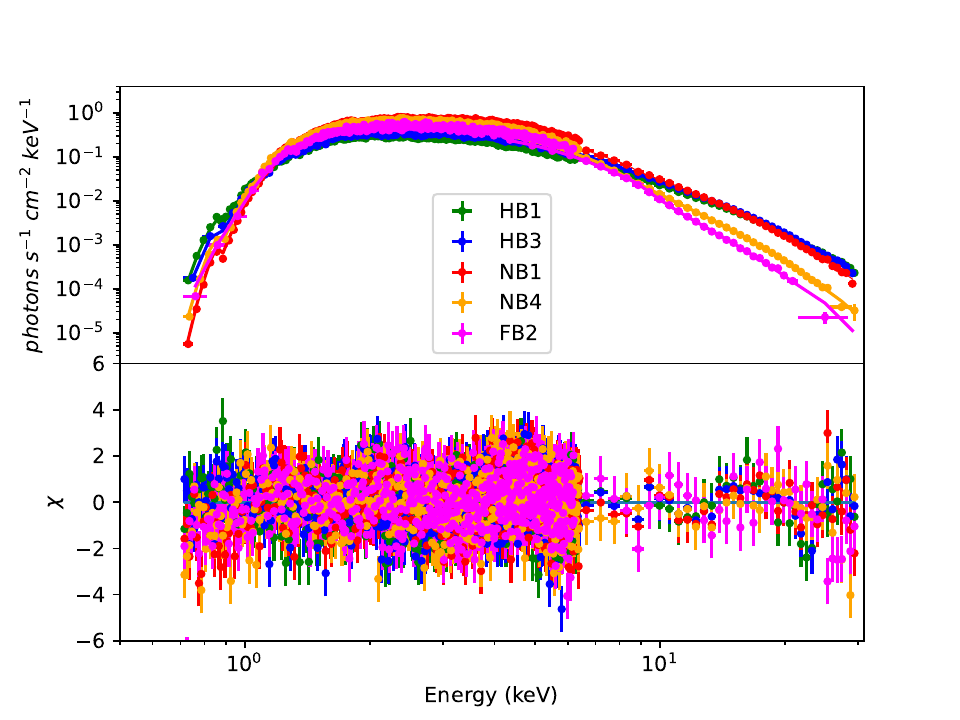} 
\caption{In the top panel of  the figure  the unfolded spectra  and the best fit model together for the segments HB1,HB3, NB1, NB4 and FB2 have been plotted. The best fit model used here is \textit{tbabs*(diskbb+Comptb)}. The residual in units of sigma for these segments is also plotted in the bottom panel of the figure.}
\label{fig:fit}
\end{figure*}

We also attempt to fit the data with \textit{Comptb} model, which takes care of both thermal and bulk Comptonization \citep{2008ApJ...680..602F}. This models is sum of two components, the soft seed photons which has not undergone up-scattering and coming directly and the component  that is Comptonized.  We fix $\gamma$ = 3 (blackbody seed photon). We find that parameter $\log A$ pegs to 8 and $\delta$ $<<1$. Since seed photons from disc  can also be Comptonized by the Corona, we included one more \textit{Comptb} component and tied the value of spectral index $\alpha$ and electron temperature $kT_e$ while fitting the data with double Comptb. For the second \textit{Comptb} component $\log A$ pegs to -8 and $\delta <<1$. For the first \textit{Comptb} component, illumination factor ($A/(A+1)$) is close to one and for the second Comptb component it is negligible. Hence, we reached to conclusion that the bulk Comptonization is not present, disk photons are seen directly and blackbody photons from the NS surface is completely Comptonized.  Hence, we fit the data with \textit{Comptb} plus \textit{diskbb} component (\textbf{Model 4}) which allow us to calculate the inner disk radius as well. We have shown the unfolded spectra using \textbf{Model 4} for the segments, HB1, HB3, NB1, NB4 and FB2 for a comparison in Figure \ref{fig:fit}.
We compared Model-4, with Model-1 and Model-2 by computing F-test chance improvement probabilities  for addition of extra parameters. \textbf{Model 3} and \textbf{Model 4} has similar reduced $\chi^2$.
 \textbf{Model 4} provides statistically better fit compared to \textbf{ Model 1} and \textbf{ Model 2} (see Table \ref{tab:stat}). We also note that \textbf{Model 2}  does not constrain the seed photon temperature in the segments  HB3-FB2. The {\tt\string cflux} command is used to derive the unabsorbed fluxes in the individual spectral components. The disk and Comptonization fluxes are computed in the energy range $0.5-50.0$~keV. Errors quoted in the best fit parameters are $1-\sigma$ errors and computed using $\Delta \chi^2=1$.
\subsection{Timing Analysis}\label{sec:timeana}
We carry out fast timing analysis using the \textit{LAXPC20} data. Using the \textit{LAXPC20} data, we generated 2-ms binned  lightcurves in the energy band $3-50$ keV for  each data points ( with 256s length) of HID. We divided these lightcurves in intervals of 4096 bins  and created the PDS for these intervals. This procedure produces PDS in the frequency range $0.12-250$ Hz for each intervals.
PDS belonging to same segments of the HID were averaged and re-binned geometrically by a factor of 1.03 in the frequency space. The PDS  were normalized  to give fractional root-mean squared power $((rms/mean)^2/Hz)$ and dead time corrected Poisson noise was subtracted \citep{1995ApJ...449..930Z, 2018MNRAS.477.5437A}. 
We performed the fitting of PDS in the frequency range $0.12-250$ Hz  using the combination of Lorentzian and power-law components. The Lorentzian function is defined as
\begin{equation}
    \frac{r^2 \Delta \nu} {2\pi [(\Delta\nu/2)^2+(\nu-\nu_c)^2]},
    \label{eq:one}
\end{equation}
where $\Delta \nu$ is the full-width at half maxima (FWHM), $\nu_c$ is the centroid frequency and $r$ is the integrated rms for the Lorentzian component \citep{2002ApJ...572..392B}.  1-$\sigma$ errors (68\% confidence) in the best fit parameters are calculated using $\Delta \chi^2=1$. We also created PDS in the energy bands $3-5$ keV, $5-8$ keV  and $8-15$ keV to carry out the energy dependent study of the observed power-spectral features.
\begin{figure}[]
\centering\includegraphics[ width=0.8\textwidth]{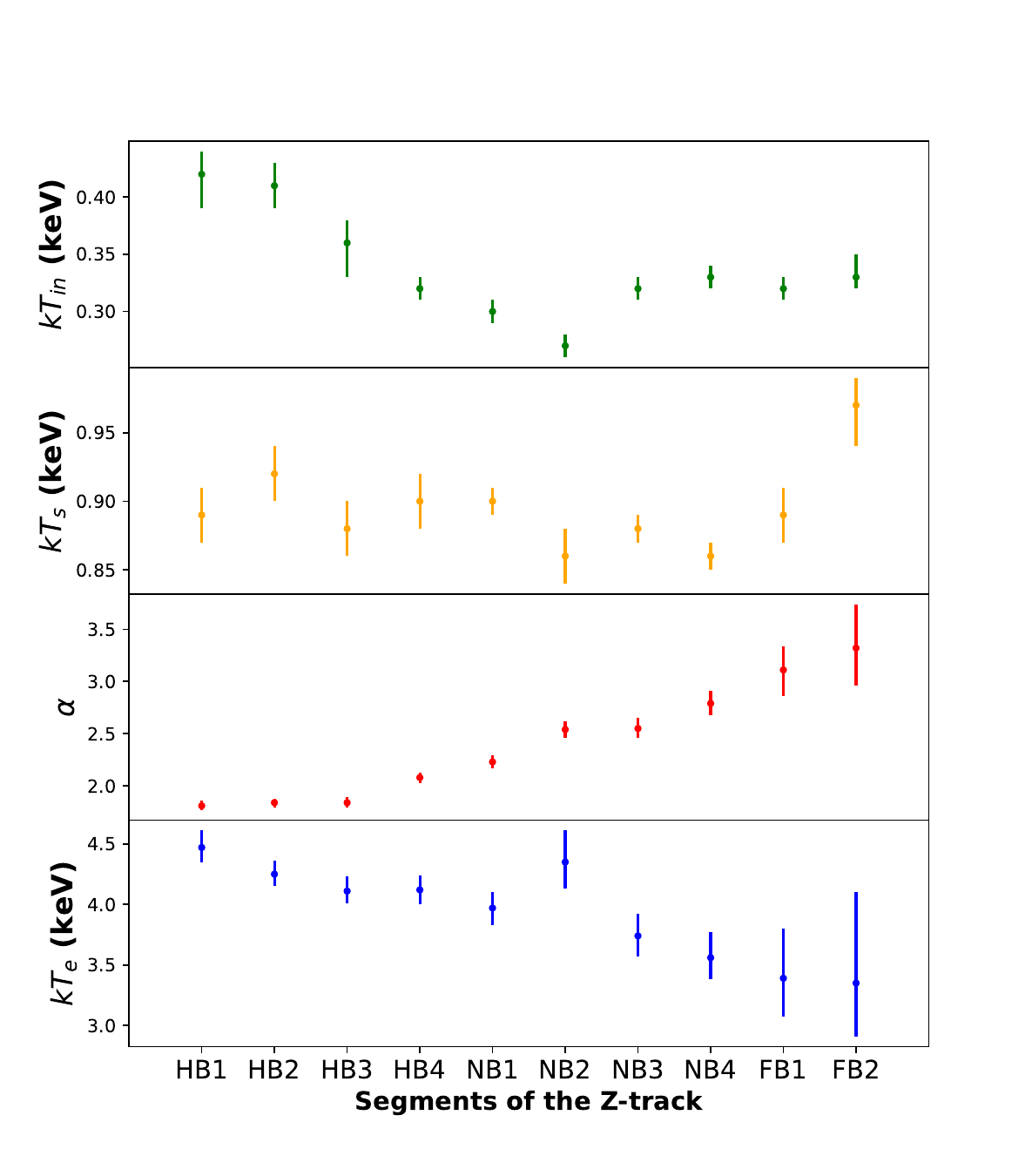}
\caption{Figure shows evolution of the parameters of \textbf{Model 4} \textit{(tbabs*(Comptb+diskbb))}. The electron temperature $kT_e$ and spectral index $\alpha$ show significant evolution as the source moves from HB to FB. The Comptonized component becomes softer from segment HB1 to FB2. Also disk temperature decreases as the source moves from the segment HB1 to NB2. For details see the text and Table \ref{tab:dbb-comptb}. \label{fig: spec-evol}}
\end{figure}
\begin{figure}[]
\centering\includegraphics[ width=0.8\textwidth]{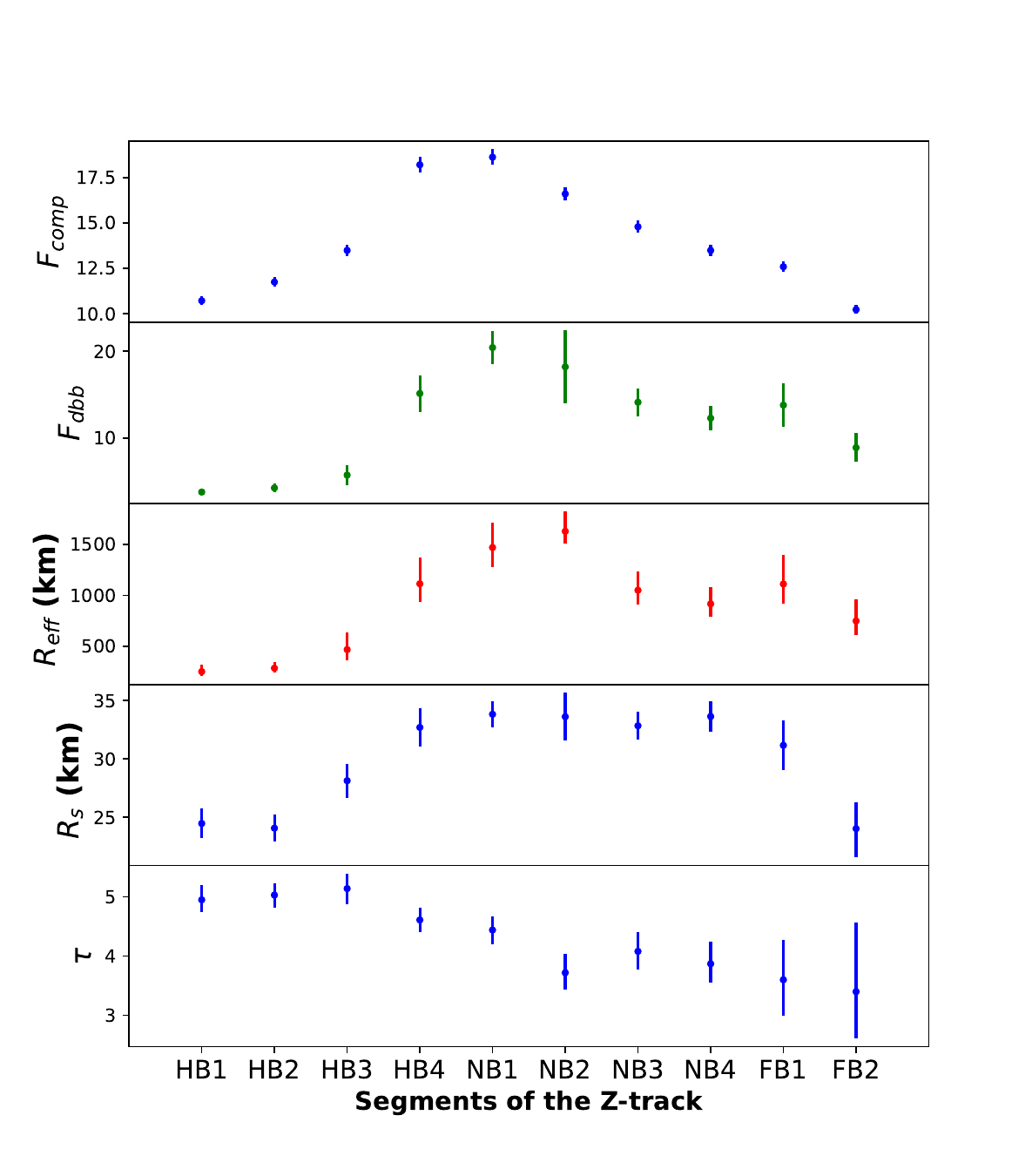}
\caption{Figure shows evolution of the Comptonized ($F_{Comp})$ flux, disk  flux ($F_{dbb})$, the effective inner disk radius ($R_{eff}$) and the optical depth ($\tau$) of the corona, calculated for \textbf{Model 4}. Significant evolution of the disk and the Comptonized flux is clearly visible. The effective disk radius and optical depth ($\tau$) also evolve significantly from the segment HB1 to FB2.   For details see the text and Table \ref{tab:flux}. \label{fig: spec-evol2}}
\end{figure}

\begin{figure*}[]
\includegraphics[width=1.0\textwidth]{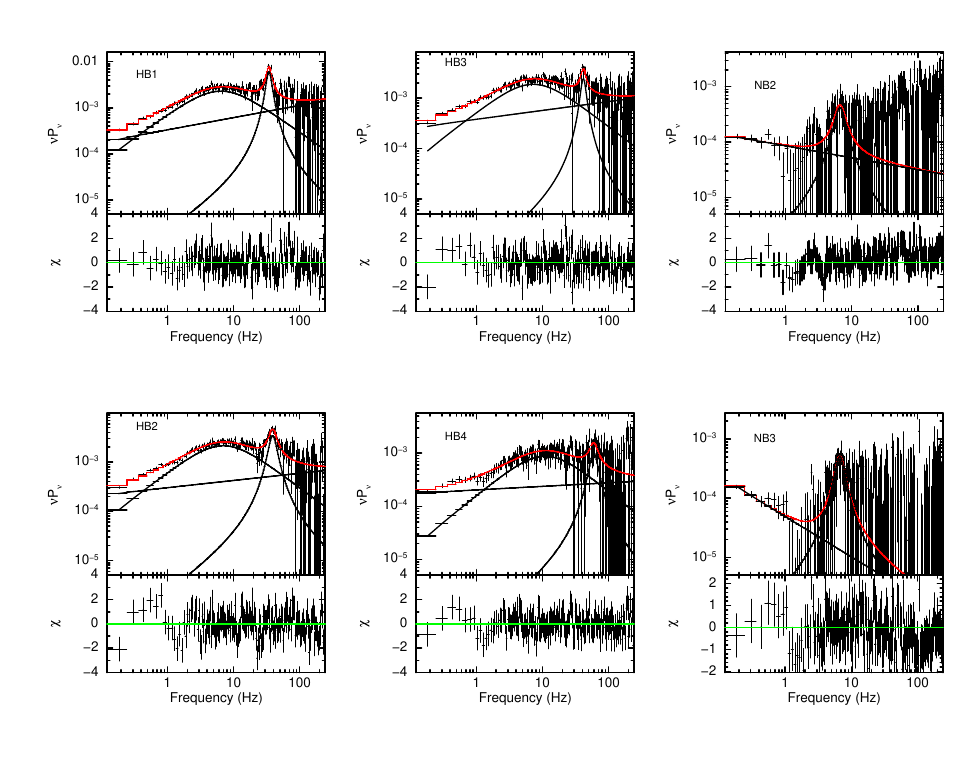}
\caption{Figure shows the PDS for HID segments HB1, HB2, HB3  HB4, NB2 and NB3 in the energy band $3-50$ keV.  In HB1, HB2 and HB3, a HBO is detected. A NBO is seen in the NB2 and NB3. The PDS for segments HB1,HB2 and HB3 are fitted with the combination of double Lorentzian and a power-law. The PDS for segments NB2 and NB3  are fitted with the combination of a power-law and a Lorentzian. Best fit model along with the observed PDS has been shown in the figure.}  \label{fig:pds}
\end{figure*}
\begin{figure}[]
\centering\includegraphics[ width=0.8\textwidth]{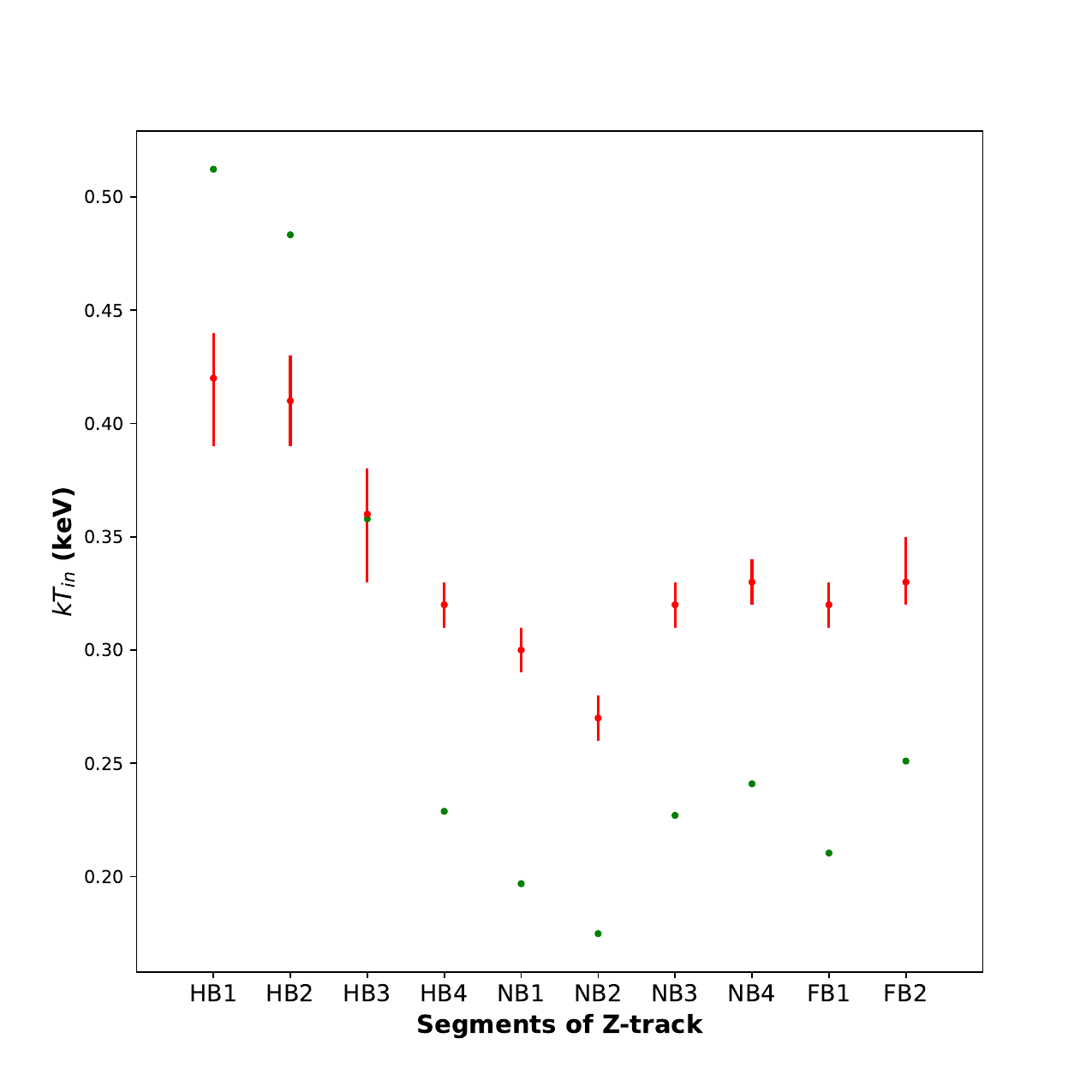}
\caption{A comparison between the observed inner disk temperature and derived values. The estimated temperature values are  denoted using green dots and the fitted values are shown using red dots with error. \label{fig: temp}}
\end{figure}
\section{Results} \label{sec:result}
\subsection{Spectral Nature of XTE J1701-462}
In Figure \ref{fig:hid}, we show the HID of the source. The HID has an extended HB and short and dipping FB. Figure \ref{fig:ccd} shows the CCD of the source. The points belonging to the HB on the CCD are clustered. The source traced similar pattern on the  CCD and HID during the Cyg-like phase of 2007 outburst \citep{2007ApJ...656..420H}.
We fit the  broad band X-ray spectra corresponding to different segments of the HID of the source  using  various popular spectral models and study  evolution of the parameters along the Z-track.
We note that both \textbf{Model 4}  and \textbf{Model 3} comprising disk-blackbody and Comptonized emission are statistically better description of the data. The seed photon energy can not be constrained for sections HB3 to FB2, 
 while using \textbf{Model 2}. It is also noted that the high blackbody temperature ($\sim$ 3 keV) obtained by fitting the broad band spectra with  \textbf{Model 1} seems to be unphysical. In  Figure \ref{fig:fit} we show the unfolded X-ray spectra for different sections fitted with \textbf{Model 4}. The best fit parameters obtained using \textbf{Model 1}, \textbf{Model 2} ,\textbf{Model 3}  and \textbf{Model 4} are listed in Table \ref{tab:bb-dbb}, Table \ref{tab:bb-comp} , Table \ref{tab:dbb-comp} and Table \ref{tab:dbb-comptb} respectively. The unabsorbed disk  and Comptonized component fluxes are listed Table \ref{tab:flux}.  The model component \textit{diskbb} has two parameters, the inner disk temperature $kT_{in}$ and the normalization $K$. The normalization $K$ is connected with the inner disk radius $R_{in}$ by the relation,
 \begin{equation}
  K=\left(\frac{R_{in}}{D_{10}}\right)^2\cos\theta,
 \end{equation}
 where  $D_{10}$ is the distance in units of 10 kpc and $\theta$ is inclination angle of the disk \citep{1984PASJ...36..741M}. We assumed distance of the source to be 8.8 kpc and inclination angle to be 60$^\circ$ \citep{2009ApJ...699...60L, 2009ApJ...696.1257L}. 
 The absence of iron line (as in the present case) or presence of weak iron line and absence of X-ray dips and eclipses suggests that inclination should be less than 70$^\circ$. Hence, we chose an average inclination angle of 60$^\circ$ to compute the inner disk radius.
 The local blackbody spectrum from the accretion disc is modified by scattering processes and  can be best represented by a diluted blackbody \citep{1995ApJ...445..780S} given by equation
 \begin{equation}
 F^{db}_\nu = \frac{1}{f^4} \pi B_\nu(fT_{eff}),
 \end{equation}
 where f is spectral hardening factor, $T_{eff}$ is effective temperature.
 The effective temperature is given by
 $T_{eff} = \frac{T_{in}}{f}$ and effective inner accretion on disc radius is given by $R_{eff} = f^2 R_{in}$.
 We computed the color correction factor $f$ using the equation \citep{2019ApJ...874...23D}
 \begin{equation}
 f \approx 1.48 + 0.33(\log \dot{m} +1) + 0.07 (\log \alpha+1) + 0.02 (\log(M/M_\odot) -1),
 \end{equation}
 here $\dot{m}$ is the mass accretion rate in unit of Eddington rate, $\alpha$ is the viscosity parameter and $M$ is the mass of neutron star.
 Color factors are listed in Table \ref{tab:flux}. $R_{eff}$ was derived  using these values and given in the Table  \ref{tab:flux}. 
 The \textit{Comptb} component has four free parameters, the spectral index $\alpha$, the electron temperature $kT_e$ of the hot corona, the seed photon energy $kT_s$, and the normalization. The radius of seed photon emitting region is given by the equation (see \citealt{1999A&A...345..100I}),
\begin{equation}
R_s = 3 \times 10^4 \sqrt{F_{comp}/(1+y)}/(kT_s)^2
\end{equation}
The above equation is obtained by equating the soft seed photon luminosity with blackbody luminosity of temperature $kT_s$
where is $y$ is the Compton y-parameter that gives relative energy gain due to Compton scattering and is given by the relation
\begin{equation}
    y=\frac{4kT_e}{m_e c^2} \tau^2,
\end{equation}
where $\tau$ is the optical depth.

The spectral index of Comptonized component is given by (see \citealt {1996MNRAS.283..193Z})
\begin{equation} 
\alpha = \left[\frac{9}{4}+\frac{1}{kT_e/m_e c^2(1+\tau/3)}\right]^{1/2}-\frac{3}{2}
\end{equation}
We solve this relation for optical depth $\tau$. The optical depth depends upon $\alpha$ and electron temperature $kT_e$.
We show the variation of the fitted (see Figure \ref{fig: spec-evol})  and derived parameters (Figure \ref{fig: spec-evol2}) of \textbf{Model 4} as a function of position on the HID. An increase in the $\alpha$ from $\sim$ 1.81 to 3.32 is observed with the movement of the source from the HB to the NB then to the FB, suggesting that the Comptonized  spectrum is becoming softer. The electron temperature $kT_e$ of the Compton corona also decreases from 4.48 to 3.35 as the source moves along the Z-track from HB to FB.  The seed photon temperature remains almost constant and remains in the narrow range $0.9-1.0$~keV. The disk temperature $kT_{in}$ decreases from $\sim$ 0.42 to $\sim$ 0.28 keV from the segment HB1 to NB2 and then remains almost unchanged in the rest of the Z-track.
Our analysis also reveals that the flux  of the Comptonized emission in the energy range $0.5-50$ keV increases from  HB1 to NB1 and then decreases from NB1 to FB2. The disk flux also show a variation similar to the Comptonized flux as the source moves along the HID. We also note that the inner disk radius $R_{eff}$ changes from $\sim$ 250 km to $\sim$ 1600 km as as the source evolves from the state HB1 to NB2 and as it moves further from NB3 to FB2 the disk radius decrease to $\sim$ 750 km.  The seed photon radius $R_s$ is found to be in the range of $\sim$ $25 - 36$~km being lowest in the upper HB (HB1 and HB2). The optical depth $\tau$ of the  corona is slightly higher ($\tau \sim 4.6-5.1)$ in the HB compared to that in NB and FB. A decrease in $\tau$ from  $\sim$ 4.6 to $\sim$ 3.4  is seen with the motion  of the source from HB4 to FB2. The $y-par$ remains almost constant in HB and then decreases along the NB and  FB (see Table \ref{tab:flux}).
\subsection{Timing behavior of XTE J1701-462}
To describe the features in the PDS we need Lorentzian and power-law (Power $\propto \nu^{-\alpha}$) components. The power-law function describes the very low frequency noise (VLFN). 
A zero centred Lorentzian describes the band limited noise (BLN).  The narrow features observed in the PDS are also represented by a Lorentzian. Following \citet{2002ApJ...572..392B}, we define the characteristic frequency of the narrow features (QPO) as,
\begin{equation}
    \nu_{char} = \sqrt{(\Delta\nu/2)^2+\nu_c^2},
\end{equation}
where $\nu_c$ and $\Delta\nu$ are centroid frequency and $FWHM$ (see Equation \ref{eq:one}).
The break frequency of BLN component is given by $\Delta \nu/2$, where $\Delta \nu$ is the FWHM of the zero centred Lorentzian component. The quality factor (Q) of the QPO is defined as $\nu_{c}/\Delta \nu$.  The best fit parameters of power-spectral features are listed in the Table \ref{tab:pds} and their rms values are given in Table \ref{tab:rms}.  We detect narrow QPOs (HBOs) with  $\nu_{char}$ (or $\nu_{HBO}$) in the range $34-40$ Hz in the HID segments HB1, HB2 and HB3 (see Figure \ref{fig:pds} and Table \ref{tab:pds}). The significance of HBOs are 12.7 $\sigma$, 10.8$\sigma$ and 5.8$\sigma$ for segments HB1, HB2 and HB3 respectively. We also detect a weak HBO-like feature in HB4 with significance 2.6$\sigma$. The significance of QPOs is computed by taking ratio of the normalization of the Lorentzian and 1-sigma negative error on normalization (see \cite{2020MNRAS.499.5891S, 2022MNRAS.512.2508M}). BLN feature with break frequency varying from $\sim$ 6.5 Hz to $\sim$ 11 Hz is seen in this region of HB (HB1-HB4). The frequency $\nu_{HBO}$ and $\nu_{break}$ increases as the source travels from HB1 to HB4. This suggest that the frequency of both components  are correlated. NBOs with $\nu_{char}$ (or  $\nu_{NBO})$  $\sim$ 6.7 Hz with rms $\sim$ 1.7\% is observed in the HID section NB2 and NB3 (see Figure \ref{fig:pds}). The significance of  NBOs is given in the Table \ref{tab:pds}. The rms of the HBOs decreases from $\sim$ 4.5\% to $\sim$2.8\%  as the source travels from HB1 to HB3. The rms of the BLN feature also shows a  decrease from $\sim$ 8.5\% to $\sim$ 5.2\% from HB1 to HB4. The total rms in the PDS decreases  as the source moves along the HID from HB1 to FB2.  We also find a broad high frequency feature with a  quality factor $<2$ with $\nu_{c}$ $\sim$ 59 Hz in  NB1. Since quality factor of this feature is small, we do not consider it as HBO. We also detect a low-frequency noise (LFN) with $\nu_{char}$ $\sim$ 9.6 Hz in this section of NB. 

We have also carried out energy dependent study of the  power-spectral features (see Table \ref{tab:erms} for details).  The strength of the HBOs increases with increase in the  energy of X-ray photons.  This trend is observed in the HID sections HB1, HB2 and HB3. However, PDS created in the  narrow energy ranges $3-5$ keV, $5-8$ keV and $8-15$ keV do not show a significant HBO for the HID segment HB4. Moreover, the rms of  BLN feature also show increasing pattern with an increase in the energy. In HB3, HBO was found to be absent in the energy range $3-5$ keV and we set an upper limit of $<1.97$\% on the rms of this feature.

\section{Discussion}\label{sec:discus}
As shown in Figure 3, both \textit{AstroSat} observations are close to the peak
of the 2022 outburst. During the \textit{AstroSat} observations the source showed
a Cyg-like behavior. Its HID has an extended HB, a short and dipping FB. The
source count rate increases by a factor of two along the HB. A similar
HID was observed in the source previously with \textit{RXTE} near the peak of the
2006 outburst \citep{2007ApJ...656..420H}. The broad-band spectral data acquired
with \textit{SXT} and \textit{LAXPC} were fitted with the four widely accepted approaches as described in the section \ref{sec:dna}. The combination of emission from a
standard accretion disk \citep{1973A&A....24..337S, 1984PASJ...36..741M} and Comptonized emission from a hot corona \citep{1996MNRAS.283..193Z, 2008ApJ...680..602F} provides a better description of the X-ray
spectra of the source. We study the evolution of the parameters of this
model along the HID. Previously, spectral evolution study along the HID
of the source in the energy band $3-100$ keV has been carried out using the \textit{RXTE} data \citep{2009ApJ...699...60L}. Hence, this is the first study covering energy band $0.7-30$ keV. The fast timing analysis revealed the presence of the HBOs ($34-40$ Hz) in the HB and the NBOs at frequency $\sim$ 6.6 Hz in the NB. We discuss the origin and nature of these features observed in the  PDS. 

The inner disk is truncated far away from the NS ($R_{eff} \sim 250-1600 $ km). Hence, the region between the inner disk rim and the magnetosphere  is filled with hot coronal plasma. A large inner accretion disk radius ($R_{eff}$) has been reported  in the Z-sources GX 340+0 \citep{2023ApJ...955..102B}  and GX 5-1 \citep{2024ApJ...977..215S}.  \citet{2023ApJ...955..102B} suggests that probably the inner accretion disk is hidden inside a large corona. Other possibility is that the disk may be truncated due to radiation pressure.  In both cases, the corona is not compact instead it is extended covering the NS surface and magnetosphere. We also note that the temperature changes by a smaller factor compared to the variation in the disk radius. Using the equation \citep{1973A&A....24..337S}
\begin{equation}
T_{in} = \left(\frac{3GM \dot{M}}{8\pi R_{in}^3\sigma}\right)^{1/4},
\end{equation}

we estimated the inner disk temperature and compared with observed temperature of the disk (see Figure \ref{fig: temp}). In this equation $M$ is mass of the NS, $\dot{M}$ is mass accretion rate.
From Figure \ref{fig: temp}, it can be seen that the trend of change in  the estimated values of $kT_{in}$ and the fitted values of $kT_{in}$ along the Z-track is similar. However, the  the estimated and observed temperatures do not exactly match, suggesting that the nature of accretion disk  may deviate from the standard accretion disk \citep{1973A&A....24..337S} at high luminosities.
The spectral modeling of emission from the source with \textit{tbabs*(diskbb+Comptb) } favor a truncated disk scenario. The disk flow probably starts deviating from Keplerian to sub-Keplerian flow at the inner edge to satisfy the inner boundary conditions at the NS surface. The transition from a Keplerian to a sub-Keplerian flow creates centrifugal barrier (CB) at the transition point where matter starts piling up vertically forming a transition layer (TL) \citep{1995ApJ...455..623C, 1997ApJ...484..313C, 1998ApJ...499..315T}.  The soft photons from the NS surface  is up-scattered by the hot electrons present in the TL \citep{2008ApJ...680..602F}.  The small illumination factor ( $A<<1$) for second Comptb  suggests that emission from the disk is  seen directly. Hence, the transition layer is geometrically thick and located  between inner edge of accretion disk ( which is truncated) and the NS magnetosphere. The photons from  the NS are  completely hidden and intercepted by an optically thick ($\tau \sim 3-5$) and geometrically thick TL. The bulk flow parameter $\delta$ is zero, suggesting that the bulk flow is suppressed due to the strong radiation pressure at the vicinity of the NS.
The bolometric unabsorbed luminosity in the energy range $0.5-50$ keV of the source increases along the HB becoming highest at the hard apex and then again decreases along the Z-track. The  disk and Comptonized luminosity follow a  similar trend. Based on the multi-frequency observations of Cyg X-2, \citet{1990A&A...235..131H} argued that accretion rate monotonically increases from HB to NB and then FB. However, opposite scenario has also been proposed to explain the `Z' tracks of Cyg-like Z-sources \citep{2011A&A...530A.102B}.  The Comptonization flux show a  systematic increase from HB1 to NB1. The location of inner edge of the disk is decided by balance between ram pressure and radiation pressure of Comptonized flux. The corona can be considered as a transition layer between inner edge of accretion disk and $R_{ISCO}$.  An increase in the radiation pressure or radiation drag may push the disk outwards. Indeed, the inner disk radius also increases from HB1 to NB1. The electron temperature of the corona shows  slight decrease (4.47 to 3.97) and the optical depth does not show much variations from HB1 to NB1. However, the Comptonization flux increases due to increase in its normalization. Since the inner rim of the disk and outer edge of the TL is moving outwards, size of the corona increases. This suggests that the corona should  become optically thin and hot as the source travels along the HB. The  increase in the soft seed photon luminosity reverse this expected change in the corona and also explains the observed brightening of the corona from HB1 to NB1.


The reduction of coronal luminosity is observed from NB1 to FB2. Also, we note that the spectra become softer (or hard color decreases) and count rate decreases along this section of  HID.  The change in the hardness ratio is basically decided by the variation in the  Comptonized component. The electron temperature decreases slightly (4.3 - 3.3 keV) and the optical depth shows slight decrease  (4.4 to 3.4) from the segment NB1 to FB2. The Sco-like Z-source GX 17+2  has also shown similar behavior in the NB \citep{2020Ap&SS.365...41A}. In GX 17+2, the optical depth decreases  and electron temperature remains almost constant along the NB. It was proposed that an increase in  the seed photon supply from the boundary-layer or the NS surface causes a small fraction of coronal material to cool down and settle down in an underlying accretion disk. This mechanism leaves slightly less denser plasma cloud, explaining observed decrease in the optical depth along the NB. However, in the present case we note that the soft seed photons from the NS surface decreases along the NB and FB. The decrease in the luminosity with an increase in the mass accretion rate can be explained by increase in anisotropy of the X-ray emitting region causing decrease in the  fraction of the X-ray flux emitted  towards the observer's line of sight. Most probably the height (H) of the corona is increasing and at the same time its shape is changing from TL with H/R$_{in}$ $<$ 1 to slightly asymmetric  flow  with $H/R_{in}$ $\sim$ 1. This may explain the decrease in the optical depth (as volume of the corona increases and hence density decreases). The increased supply of  seed photons from the NS surface,  anticipated due to increase in accretion rate can cause decrease in the coronal temperature from NB1 to FB2.  A decrease in the polarization degree (PD) has been observed as the source moves HB to  NB. The decrease in the scattering efficiency of the corona from from the hard apex to the lower NB can explain the observed decrease in the PD.

We observe a dipping FB in this source, such a dipping behavior has been observed in Z-sources,  GX 340+0 \citep{1998ApJ...499L.191J}, GX 5-1 \citep{1998ApJ...504L..35W} and Cyg X-2 \citep{2018MNRAS.474.2064M}. This source also has shown a dipping FB during its previous outburst \citep{2007ApJ...656..420H}. The dipping FB in the source is associated with a reduced optical depth and a lower coronal temperature.  This is  opposite to behavior observed in Cyg X-2 where the optical depth was found to increase during the X-ray dips \citep{2018MNRAS.474.2064M}. Multi-wavelength study of Cyg X-2 suggested that the X-ray dip is caused by absorption of an extended coronal emission by the structure in the outer accretion disk \citep{2011A&A...530A.102B}. Hence, changing corona geometry (X-ray emitting region becomes  anisotropic in the  FB) can explain the X-ray dips observed in the source. 

This source has shown, HBOs, NBOs and a pair of kHz QPOs during the previous outburst \citep{2007ApJ...656..420H}. The source exhibited kHz QPO pairs in Z-phase and single kHz QPO in atoll-phase \citep{2010MNRAS.408..622S}.  In the Z-phase, kHz QPOs were weaker and broader compared to the atoll-phase. The timing properties of the source was studied in detail  during the Cyg-like phase \citep{2007ApJ...656..420H} where the source exhibited dipping FB.  It was noted that frequency of HBO increases and  rms strength decreases along the HB.  During \textit{AstroSat} observation also the source was in Cyg-like phase and showed dipping FB, suggesting that temporal and spectral properties of the source should be similar.  During the AstroSat observations the source exhibited  HBOs in the frequency range $34-40$ Hz. The strength of the  HBOs observed in the source  decreases as the source travels from HB1 to HB3 and then they disappear as the source further moves down the Z-track.  The Comptonization flux increases from HB1 to HB4. However ratio of the Comptonization flux and total flux decreases  along the HB and hence the HBO rms is correlated with the percentage contribution of Comptonized emission to the total flux.  \citet{2015ApJ...799....2B} investigated the HBO properties of the source using all available  \textit{RXTE} observations of the source during its 2006 outburst. They also found that the  HBO rms decreases from top-left of the HB to the hard-apex.
The rms of the HBOs and the BLN component increases with increase in the photon energy from 4 to 12 keV (see Table 7). Energy dependent studies of HBOs of this source using data from \textit{RXTE} satellite also provided a similar result in this source \citep{2015ApJ...799....2B}. GX 340+0 also shows increase in the HBO strength with increasing photon energy \citep{2000ApJ...537..374J}. Hard nature of the HBOs and the BLN component suggests that their origin is linked with the hot corona around the compact object. We find a positive correlation between the break frequency and the HBO frequency, consistent with the previous observation of the source \citep{2015ApJ...799....2B}.  Moreover, frequency resolved spectroscopy of the HBOs suggests that these oscillations are produced in the Comptonizing region \citep{2006A&A...453..253R}. 

In many  models of QPOs, the  kHz QPOs and HBOs are linked with the inner disk radius \citep{1998ApJ...508..791M, 1999PhRvL..82...17S}.  According to these models the upper kHz QPOs depends upon the location of inner edge of accretion disk. The observed kHz QPOs \citep{2010MNRAS.408..622S} in the XTE J1701-462 gives $R_{in} \sim 20$ km. In the present case, disk is truncated away from the NS surface. If spectral nature of the source  during the Z-phase of the previous outburst and the present outburst are similar then the origin of kHz QPOs can not be explained in the framework of these QPO models. Presently, we have detected only HBOs and NBOs in the Cyg-like Z-phase of the source. 

The Lense-Thirring (LT) precession model has been proposed to explain the HBOs observed in NS-LMXBs and LFQPOs in black hole binaries \citep{2009MNRAS.397L.101I}. According to this model precession of radially extended hot central corona  produces the observed QPOs in these two class of objects. The outer radius of the hot flow is decided by the truncation radius of the disk. The disk is truncated $\sim$ 250-1100 km in the segments HB1 to HB4 sections of the Z-track.  We take outer radius $r_o = R_{eff}/R_g$ and inner radius $r_i = R_{NS}/R_g$ of the hot radial flow.  The LT precession frequency is computed using the formula \citep{2009MNRAS.397L.101I}
\begin{equation}
 \nu_p = \frac{5}{\pi} \frac{a[1-(r_i/r_o)^{1/2}]}{r_o^{5/2} r_i^{1/2} [1-(r_i/r_o)^{5/2}]} \frac{c}{R_g},
 \end{equation}
 where $R_g = 2.1$ km is gravitational radius of 1.4 $M_\odot$ neutron star. $R_{NS} = 10$ km is NS radius. 
The above equation gives the QPO frequency of 0.2 Hz for $R_{eff} =250$ km. Hence, this model can not explain the origin of the HBOs observed in the NS-LMXBs. \citet{2014ApJ...786..119L} observed  a positive correlation between the inner disk radius and the HBO frequency. We also observed a similar correlation,  which is hard to explain in the framework of the LT precession model. 
 \textbf{Probably}, the LT precesses  diferentially with a higher precession frequency at the inner region of TL  compared to outer region \citep{2016MNRAS.458.3655V, 2018ApJ...866..122H}. Here, we can assume that TL is slightly warped where  height of TL is higher in the inner region compared to that in outer region. Hence more modulation is produced by the faster precessing inner region.  

Recently phase resolved polarimetry of a strong LFQPOs,  seen in the transient blackhole candidate Swift J1727.8–1613, has been carried out \citep{2024ApJ...961L..42Z}. No modulation in the polarization degree and the polarization angle is observed with the QPO phase. Hence polarimetric observations are also not in favor of the LT precession model. LT precession model of jet and inner disk ring has been used to explain the LFQPOs seen in Swift J1727.8-1613 \citep{2024MNRAS.529.4624Y}. LFQPO in the hard X-rays  was detected in Swift J1727.8-1613 with \textit{AstroSat} \citep{2024MNRAS.531.1149N}. It was suggested that the oscillations in a hot and dense downstream flow at the vicinity of the blackhole produces the QPOs \citep{2024MNRAS.531.1149N}.

IGR 17480-2446, which is a  transient mili-second pulsars with 11 Hz pulse period, exhibited HBOs ($35 - 50$ Hz) and kHz QPOs \citep{2012ApJ...753...84B, 2012ApJ...759L..20A}. This source also showed transition from atoll to luminous Z-phase \citep{2011MNRAS.418..490C}. To explain the nature of HBO in this source the \citet{2012ApJ...759L..20A} suggested that precession due to frame dragging alone is not sufficient to explain the observed frequency of HBOs. They proposed that a warp is induced by magnetic field in the hot inner flow  and that precesses with frequency $\nu_{mag}$. The magnetic precession depends on the magnetic moment $\mu$. The net precession frequency is combination of magnetic and frame dragging effects.

As HBOs are linked to the Comptonizing component,  other kind of oscillations in the corona can produce these frequencies. \citet{2010MNRAS.404..738C} proposed the oscillating hot corona model to explain the low frequency QPOs seen in X-ray binaries. They suggested that a magneto-acoustic wave propagating in the hot corona can produce resonant peaks in the PDS. 

Another popular  model proposed to explain the origin of the QPOs is two-oscillator model \citep{1999ApJ...522L.113O}. According to this model hot, large and inhomogeneous blobs are formed at the outer edge of the TL and injected into the NS magnetosphere. If the blobs formed at the outer edge of the TL are rotating differentially, having higher angular velocity at inner edge compared to outer edge. These blobs with certain velocity distributions are injected to the NS magnetosphere from the inner edge of TL. Such blobs under influence of Coriolis force behave as Keplerian oscillators  with two modes : radial and vertical \citep{1999ApJ...522L.113O}. The radial mode is identified with upper kHz QPO and the mode perpendicular to the accretion disk produces the HBOs. The lower kHz QPO is identified with the Keplerian frequency of the blobs at the inner edge of TL. These hot  plasma balls intercept the photons from the NS surface and up-scatter them.  Hence, we expect  rms strength of the HBOs to be positively correlated with the energy of photons. A positive correlation seen between the HBO-rms and the photon energy support this scenario. The radiation pressure increases as the source moves from HB1 to HB4  due to increase in the source luminosity. The increasing radiation pressure disrupts the flow of hot blobs responsible for HBOs. Therefore, HBOs become weaker as the source moves down the HB and disappear as source moves  further   along the Z-track. The observed LFQPOs (HBOs and NBOs) can also be explained in the framework of  oscillating shock wave model \citep{1996ApJ...457..805M}. In this model,  shock oscillates radially with a time period equal to the cooling time scale of the corona. NBOs at frequency $\sim$ 6.6 Hz are seen in the middle and lower NB. The frequency of NBOs always lies in the narrow range ($\sim 5-7 $ Hz), suggesting that they are linked with some kind of fundamental frequency of accretion flow close to the NS \citep{1987A&A...186..153H}. According model proposed by \citet{1987A&A...186..153H},  at high accretion rate the inomogeneous plasma is  accreted at the free fall velocity and causes density fluctuations in a layer above the NS surface. The density fluctuations propagates with local sound speed on the NS surface given by 
\begin{equation}
v_s = 4.2 \times 10^7 R_6^{-1/4} \left(\frac{M}{M_\odot} \times \frac{L}{L_{Edd}} \right), cm ~ s^{-1}
\end{equation}
where $R_6$ is NS radius in units of $10^6$ cm \citep{1987A&A...186..153H}. The  propagation of these fluctuations on the NS surface produces NBOs. Due to variations in  mass accretion rate, a spectrum of density fluctuations will be produced, the fundamental tone of which is given by
\begin{equation}
\nu_s = \frac{v_s}{2\pi R} Hz,
\end{equation}
where R is the neutron star radius. It is clear that the $\nu_s$ depends upon the NS radius, mass accretion rate (which is always close to Eddington limit for 1.4 solar mass NS) and mass of the NS \citep{1987A&A...186..153H}. Therefore, the NBO frequency lies in a  narrow range around 6 Hz. Alternatively, it can be produced by acoustic oscillations of a spherical shell around the NS \citep{2001ApJ...555L..45T}.
In this scenario, transition layer can become spherical at high accretion rate. The spherical shell oscillates with a frequency,
\begin{equation}
\nu_s = \frac{fv_s}{L} Hz,
\end{equation}
where, f=0.5 or $1/2\pi$, depending upon the boundary conditions and $L$ is size of the spherical shell. In this model of NBO,  the oscillation frequency depends upon size of the spherical shell which can vary from source to source and hence suffers difficulties. Alternatively the  observed NBOs can also be explained in the framework of oscillating corona model \citep{2010MNRAS.404..738C}.

\begin{table*}[ht]
\caption{Best fit parameters obtained by fitting the spectra for all  segments of HID using model \textit{tbabs*(bbodyrad+diskbb)} (\textbf{Model 1}). The fit parameters  are $N_H$ in units of $10^{22} cm^{-2}$, , disk temperature $kT_{in}$, disk normalization $N_{MCD}$, blackbody temperature $kT_{BB}$ and blackbody normalization $N_{BB}$. The reduced $\chi^2$ ($\chi^2/dof)$ is also given in the last column.}
    \begin{tabular}{lllllll}
    \hline
    HID-Sections& \multicolumn{6}{c}{Best fit parameters}\\
    \hline
    \hline
    $-$ & $N_H$ & $kT_{in}$ (keV)  & $N_{MCD}$ & $kT_{BB}$ (keV) & $N_{BB}$ & $\chi^2/dof$\\
    \hline
    HB1 & 1.71$^{+0.06}_{-0.06}$ & 1.84$^{+0.03}_{-0.04}$ & 46.12$^{+3.80}_{-4.27}$ & 2.96$^{+0.11}_{-0.11}$ & 2.84$^{+0.75}_{-0.51}$ & 686/633 \\
HB2 & 1.73$^{+0.05}_{-0.07}$ & 1.89$^{+0.04}_{-0.05}$ & 47.54$^{+4.15}_{-5.17}$ & 2.94$^{+0.11}_{-0.12}$ & 3.41$^{+0.98}_{-0.69}$ & 697/633\\
HB3 & 1.79$^{+0.02}_{-0.02}$ & 1.92$^{+0.04}_{-0.04}$ & 51.61$^{+4.65}_{-4.12}$ & 3.03$^{+0.11}_{-0.11}$ & 2.75$^{+0.89}_{-0.60}$ & 720/633\\
HB4 & 2.10$^{+0.02}_{-0.02}$ & 1.95$^{+0.04}_{-0.04}$ & 73.92$^{+5.75}_{-5.33}$ & 3.22$^{+0.06}_{-0.05}$ & 2.12$^{+0.35}_{-0.29}$ & 754/634\\
NB1 & 2.08$^{+0.02}_{-0.02}$ & 1.92$^{+0.02}_{-0.02}$ & 80.63$^{+4.28}_{--4.10}$ & 3.11$^{+0.07}_{-0.06}$ & 2.06$^{+0.38}_{-0.31}$ & 930/634\\
NB2 & 2.07$^{+0.03}_{-0.02}$ & 1.78$^{+0.02}_{-0.02}$ & 99.28$^{+6.54}_{-6.01}$ & 3.00$^{+0.06}_{-0.06}$ & 1.80$^{+0.27}_{-0.23}$ & 810/634 \\
NB3 & 2.01$^{+0.02}_{-0.02}$ & 1.71$^{+0.02}_{-0.02}$ & 109.48$^{+6.15}_{-5.72}$ & 2.76$^{+0.06}_{-0.06}$ & 2.29$^{+0.53}_{-0.35}$& 821/634\\
NB4 & 2.00$^{+0.02}_{-0.02}$ & 1.59$^{+0.02}_{-0.02}$ & 139.34$^{+7.36}_{-6.88}$ & 2.58$^{+0.06}_{-0.05}$ & 2.49$^{+0.68}_{-0.51}$ & 814/634\\
FB1 & 2.07$^{+0.03}_{-0.02}$ & 1.61$^{+0.03}_{-0.02}$ & 135.05$^{+8.87}_{-9.55}$ & 2.64$^{+0.12}_{-0.08}$ & 1.30$^{+0.39}_{-0.25}$ & 725/634\\
FB2 & 2.02$^{+0.02}_{-0.02}$ & 1.70$^{+0.01}_{-0.02}$ & 83.54$^{+5.75}_{-5.35}$ & 2.64$^{+0.09}_{-0.13}$ & 0.98$^{+0.64}_{-0.37}$ & 837/634\\
      \hline
    \end{tabular}
    
    \label{tab:bb-dbb}
\end{table*}

\begin{table*}[ht]
\caption{Best fit parameters obtained by fitting the spectra for  different segments of HID using model \textit{tbabs*(bbodyrad+nthComp)} (\textbf{Model 2}). The fit parameters  are the hydrogen column density $N_H$ in units of $10^{22} cm^{-2}$, the photon index $\Gamma$, the electron temperature $kT_e$, the seed photon temperature $kT_s$ and normalization of the Comptonized component $N_{COMP}$ the blackbody temperature $kT_{BB}$ and the blackbody normalization $N_{BB}$. The reduced $\chi^2$ ( $\chi^2/dof$ ) is also presented in the last column of the table.}
\begin{tabular}{lllllllll}
    \hline
    HID-Sections& \multicolumn{8}{c}{Best fit parameters}\\
    \hline
    \hline
$-$ & $N_H$ & $\Gamma$ & $kT_e$ (keV) & $kT_s$ (keV) & $N_{COMP}$ & $kT_{bb}$ (keV) & $N_{BB}$ & $\chi^2/dof$ \\
\hline
HB1 & 1,85$^{+0.07}_{-0.07}$ & 1.98$^{+0.02}_{-0.01}$ & 3.74$^{+0.04}_{-0.03}$ & 0.31$^{+0.10}_{-0.08}$ & 1.81$^{+0.39}_{-0.24}$ & 1.21$^{+0.02}_{-0.01}$ & 194.35$^{+11.66}_{-15.02}$ & 696/634\\
HB2 & 1.93$^{+0.06}_{-0.05}$ & 1.98$^{+0.02}_{-0.01}$ & 3.60$^{+0.03}_{-0.04}$ & 0.27$^{+0.06}_{-0.08}$ & 2.07$^{+0.29}_{-0.25}$ & 1.24$^{+0.01}_{-0.01}$ & 208.35$^{+14.37}_{-13.65}$ & 690/634\\
HB3 & 1.98$^{+0.06}_{-0.13}$ & 1.94$^{+0.03}_{-0.02}$ & 3.47$^{+0.05}_{-0.04}$ & 0.15 (fix)  & 2.20$^{+0.17}_{-0.17}$ & 1.19$^{+0.02}_{-0.02}$ & 290.76$^{+26.30}_{-21.14}$ & 710/635\\
HB4 & 2.26$^{+0.05}_{-0.05}$ & 1.97$^{+0.02}_{-0.02}$ & 3.34$^{+0.04}_{-0.04}$ & 0.21 (fix) & 2.91$^{+0.24}_{-0.23}$ & 1.17$^{+0.01}_{-0.01}$ & 506.54$^{+41.56}_{-37.85}$ & 713/635 \\
NB1 & 2.28$^{+0.03}_{-0.03}$ & 2.04$^{+0.02}_{-0.02}$ & 3.25$^{+0.03}_{-0.04}$ & 0.20 (fix) & 3.23$^{+0.15}_{-0.18}$ & 1.16$^{+0.01}_{-0.01}$ & 532.16$^{+31.16}_{-34.35}$ & 840/635\\
NB2 & 2.20$^{+0.05}_{-0.04}$ & 2.10$^{+0.03}_{-0.04}$ & 3.16$^{+0.05}_{-0.05}$ & 0.19 (fix) & 2.94$^{+0.39}_{-0.37}$ & 1.09$^{+0.04}_{-0.04}$ & 717.26$^{+55.84}_{-51.47}$ & 771/635 \\
NB3 & 2.21$^{+0.05}_{-0.04}$ & 2.20$^{+0.06}_{-0.07}$ & 3.08$^{+0.03}_{-0.05}$ & 0.20 (fix) & 3.16$^{+0.22}_{-0.22}$ & 1.08$^{+0.01}_{-0.01}$ & 691.48$^{+41.88}_{-38.02}$ & 754/635\\
NB4 & 2.26$^{+0.04}_{-0.04}$ & 2.36$^{+0.03}_{-0.03}$ & 2.99$^{+0.04}_{-0.04}$ & 0.20 (fix) & 3.65$^{+0.11}_{-0.23}$ & 1.04$^{+0.01}_{-0.01}$ & 766.22$^{+43.54}_{-41.99}$ & 761/635 \\
FB1 & 2.38$^{+0.07}_{-0.06}$ & 2.42$^{+0.05}_{-0.05}$ & 2.86$^{+0.08}_{-0.08}$ & 0.21 (fix) & 3.33$^{+0.38}_{-0.39}$ & 1.05$^{+0.01}_{-0.01}$ & 775.05$^{+48.17}_{-45.39}$ & 684/635\\
FB2& 2.21$^{+0.07}_{-0.07}$ & 2.45$^{+0.06}_{-0.06}$ & 2.86$^{+0.19}_{-0.16}$ & 0.21 (fix) & 2.85$^{+0.33}_{-0.26}$ & 1.17$^{+0.01}_{-0.01}$ & 385.48$^{+19.69}_{-18.18}$ & 830/635\\
 \hline
    \end{tabular}
    \label{tab:bb-comp}
\end{table*}
\begin{table*} 
\caption{Best fit parameters obtained by fitting the spectra of different sections of the Z-track using model \textit{tbabs(nthComp+diskbb)} (\textbf{Model 3}). The parameters of the fits are $N_H$ in units of $10^{22} cm^{-2}$, photon index $\Gamma$, electron temperature $ kT_e$, disk temperature $kT_{in}$, normalization of Comptonized component  $N_{comp}$ , seed photon temperature $kT_{s}$, disk normalization $N_{MCD}$.}
    \begin{tabular}{lllllllll}
    \hline
    HID-Sections & \multicolumn{8}{c}{Parameters}\\
    \hline
    \hline
  $-$  & $N_H$ & $\Gamma$ & $kT_e$ (keV) &  $kT_{s}$ (keV) & $N_{COMP}$ & $kT_{in}$ (keV) & $N_{MCD}$ ($\times 10^4$) &$\chi^2/dof$ \\ 
  \hline
HB1 & 1.94$^{+0.07}_{-0.07}$ & 2.73$^{+0.03}_{-0.04}$ & 4.82$^{+0.11}_{-0.15}$ & 0.87$^{+0.01}_{-0.02}$ & 0.51$^{+0.03}_{-0.03}$ & 0.41$^{+0.02}_{-0.02}$ & 0.90$^{+0.44}_{-0.29}$ & 660/632\\
HB2 & 2.05$^{+0.07}_{-0.06}$ & 2.77$^{+0.03}_{-0.01}$ & 4.59$^{+0.13}_{-0.13}$ & 0.89$^{+0.02}_{-0.02}$ & 0.54$^{+0.02}_{-0.03}$ & 0.41$^{+0.02}_{-0.02}$ & 1.11$^{+0.53}_{-0.35}$ & 663/632\\
HB3 & 2.25$^{+0.08}_{-0.07}$ & 2.78$^{+0.03}_{-0.02}$ & 4.45$^{+0.16}_{-0.14}$ & 0.88$^{+0.02}_{-0.02}$ & 0.66$^{+0.04}_{-0.04}$ & 0.36$^{+0.02}_{-0.02}$ & 2.74$^{+1.70}_{-1.14}$ & 686/632\\
HB4 & 2.83$^{+0.11}_{-0.11}$ & 2.99$^{+0.04}_{-0.05}$ & 4.43$^{+0.18}_{-0.16}$ & 0.86$^{+0.02}_{-0.02}$ & 1.06$^{+0.05}_{-0.05}$ & 0.32$^{+0.01}_{-0.01}$ & 12.53$^{+4.44}_{-3.21}$ & 666/632\\
NB1 & 2.90$^{+0.06}_{-0.07}$ & 3.13$^{+0.05}_{-0.06}$ & 4.29$^{+0.14}_{-0.16}$ & 0.87$^{+0.01}_{-0.01}$ & 1.09$^{+0.04}_{-0.05}$ & 0.30$^{+0.01}_{-0.01}$ & 20.10$^{+4.95}_{-4.25}$ & 712/632\\
NB2 & 2.92$^{+0.07}_{-0.05}$ & 3.45$^{+0.06}_{-0.06}$ & 4.76$^{+0.28}_{-0.29}$ & 0.82$^{+0.01}_{-0.01}$ & 1.26$^{+0.07}_{-0.08}$ & 0.29$^{+0.01}_{-0.01}$ & 27.74$^{+5.55}_{-3.69}$ & 732/632\\
NB3 & 2.74$^{+0.07}_{-0.06}$ & 3.50$^{+0.06}_{-0.07}$ & 4.22$^{+0.17}_{-0.29}$ & 0.85$^{+0.02}_{-0.01}$ & 1.03$^{+0.06}_{-0.05}$ & 0.32$^{+0.01}_{-0.01}$ & 12.23$^{+4.85}_{-2.20}$ & 698/632\\
NB4 & 2.79$^{+0.06}_{-0.06}$ & 3.84$^{+0.08}_{-0.08}$ & 4.31$^{+0.27}_{-0.31}$ & 0.84$^{+0.01}_{-0.01}$ & 1.08$^{+0.06}_{-0.06}$ & 0.30$^{+0.01}_{-0.01}$ & 16.92$^{+3.59}_{-3.48}$ & 709/632\\
FB1 & 2.78$^{+0.11}_{-0.10}$ & 3.75$^{+0.20}_{-0.22}$ & 3.41$^{+0.44}_{-0.30}$ & 0.85$^{+0.02}_{-0.02}$ & 1.01$^{+0.07}_{-0.07}$ & 0.31$^{+0.02}_{-0.02}$ & 11.53$^{+6.98}_{-4.31}$ & 654/632\\
FB2 & 2.63$^{+0.15}_{-0.12}$ & 4.14$^{+0.28}_{-0.39}$ & 3.58$^{+0.58}_{-0.39}$ & 0.94$^{+0.02}_{-0.02}$ & 0.63$^{+0.04}_{-0.04}$ & 0.34$^{+0.03}_{-0.02}$ & 6.13$^{+3.18}_{-1.53}$ & 806/632\\
\hline
\end{tabular}
    \label{tab:dbb-comp}
\end{table*}
\begin{table*} 
\caption{Best fit parameters obtained by fitting the spectra of different sections of the Z-track using model \textit{tbabs(Comptb+diskbb)} (\textbf{Model 4}). The parameters of the fits are $N_H$ in units of $10^{22} cm^{-2}$, energy index $\alpha$, electron temperature $ kT_e$, disk temperature $kT_{in}$, normalization of Comptonized component  $N_{comp}$ , seed photon temperature $kT_{s}$, disk normalization $N_{MCD}$.}
    \begin{tabular}{lllllllll}
    \hline
    HID-Sections & \multicolumn{8}{c}{Parameters}\\
    \hline
    \hline
  $-$  & $N_H$ & $\alpha$ & $kT_e$ (keV) &  $kT_{s}$ (keV) & $N_{COMP}$ & $kT_{in}$ (keV) & $N_{MCD}$ ($\times 10^4$) &$\chi^2/dof$ \\ 
  \hline
HB1 & 1.94$^{+0.07}_{-0.07}$ & 1.81$^{+0.05}_{-0.04}$ & 4.47$^{+0.13}_{-0.12}$ & 0.89$^{+0.02}_{-0.02}$ & 0.10$^{+0.01}_{-0.01}$ & 0.42$^{+0.02}_{-0.03}$ & 0.87$^{+0.45}_{-0.34}$ & 661/632\\
HB2 & 2.04$^{+0.07}_{-0.07}$ & 1.84$^{+0.04}_{-0.05}$ & 4.25$^{+0.11}_{-0.10}$ & 0.92$^{+0.02}_{-0.02}$ & 0.12$^{+0.01}_{-0.01}$ & 0.40$^{+0.02}_{-0.02}$ & 1.08$^{+0.48}_{-0.34}$ & 664/632\\
HB3 & 2.25$^{+0.12}_{-0.11}$ & 1.85$^{+0.04}_{-0.05}$ & 4.11$^{+0.12}_{-0.13}$ & 0.88$^{+0.02}_{-0.02}$ & 0.13$^{+0.01}_{-0.02}$ & 0.36$^{+0.02}_{-0.03}$ & 2.73$^{+1.91}_{-1.18}$ & 685/632\\
HB4 & 2.92$^{+0.1}_{-0.1}$ & 2.08$^{+0.05}_{-0.05}$ & 4.12$^{+0.12}_{-0.12}$ & 0.90$^{+0.01}_{-0.01}$ & 0.20$^{+0.01}_{-0.01}$ & 0.32$^{+0.01}_{-0.01}$ & 13.02$^{+6.05}_{-4.22}$ & 676/632\\
NB1 & 3.05$^{+0.07}_{-0.07}$ & 2.23$^{+0.06}_{-0.06}$ & 3.97$^{+0.13}_{-0.14}$ & 0.90$^{+0.01}_{-0.01}$ & 0.21$^{+0.02}_{-0.01}$ & 0.30$^{+0.01}_{-0.01}$ & 21.62$^{+6.10}_{-4.81}$ & 717/632\\
NB2 & 2.98$^{+0.15}_{-0.15}$ & 2.54$^{+0.08}_{-0.08}$ & $4.35^{+0.26}_{-0.22}$ & 0.86$^{+0.02}_{-0.01}$ & 0.19$^{+0.01}_{-0.01}$ & 0.28$^{+0.01}_{-0.01}$ & 27.48$^{+5.64}_{-3.59}$ & 735/632\\
NB3 & 2.79$^{+0.08}_{-0.09}$ & 2.55$^{+0.09}_{-0.08}$ & 3.74$^{+0.17}_{-0.16}$ & 0.88$^{+0.01}_{-0.01}$ & 0.17$^{+0.01}_{-0.01}$ & 0.32$^{+0.01}_{-0.01}$ & 12.09$^{+4.21}_{-3.22}$ & 704/632\\
NB4 & 2.75$^{+0.07}_{-0.07}$ & 2.79$^{+0.10}_{-0.11}$ & 3.56$^{+0.21}_{-0.18}$ & 0.86$^{+0.01}_{-0.01}$ & 0.16$^{+0.01}_{-0.01}$ & 0.33$^{+0.01}_{-0.01}$ & 9.56$^{+3.44}_{-2.57}$ & 712/632\\
FB1 & 2.87$^{+0.11}_{-0.12}$ & 3.11$^{+0.23}_{-0.25}$ & 3.39$^{+0.41}_{-0.29}$ & 0.89$^{+0.02}_{-0.02}$ & 0.16$^{+0.01}_{-0.01}$ & 0.32$^{+0.02}_{-0.02}$ & 13.94$^{+7.22}_{-4.81}$ & 658/632\\
FB2 & 2.69$^{+0.12}_{-0.11}$ & 3.32$^{+0.32}_{-0.36}$ & 3.35$^{+0.57}_{-0.44}$ & 0.97$^{+0.02}_{-0.01}$ & 0.13$^{+0.01}_{-0.01}$ & 0.34$^{+0.03}_{-0.02}$ & 7.03$^{+3.94}_{-2.57}$ & 808/632\\
\hline
\end{tabular}
    \label{tab:dbb-comptb}
\end{table*}
\begin{table*}[ht]
 \caption{The table provides the Comptonization flux $F_{Comp}$ and disk flux $F_{dbb}$ in the energy range $0.5-50.0$ keV. All the fluxes are in units of $10^{-9}$ ergs/s/cm$^2$. The disk radius $R_{eff}$, seed photon radius $R_s$, optical depth $\tau$ and color correction factor are also given. See text for details. The derived values are for the \textbf{Model 4} }
    \begin{tabular}{lllllllllll}
    \hline
    HID-Sections& \multicolumn{7}{c}{Derived fluxes and parameters}\\
    \hline
    \hline
-- &  $F_{Comp}$ & $F_{dbb}$  & $R_{eff} (km)$ & $R_{in} (km)$ & $R_{s}$ (km) & $\tau$ & y-par &f\\
\hline
HB1 &    10.72$\pm$0.25 & 3.80$\pm$0.28 & 252.86$^{+69.54}_{-47.58}$ & 82.22$^{+21.40}_{-14.58}$ &24.46$^{+1.28}_{-1.28}$ & 4.95$^{+0.25}_{-0.21}$& 0.85$\pm$0.08 & 1.75\\
HB2& 11.76$\pm$0.27 & 4.27$\pm$0.49 & 286.55$^{+66.29}_{-46.81}$ & 91.81$^{+20.46}_{-14.45}$ & 24.06$^{+1.18}_{-1.18}$ & 5.03$^{+0.20}_{-0.22}$ & 0.84$\pm$0.07 & 1.77\\
HB3 & 13.49$\pm$0.31 & 5.75$\pm$1.09 & 468.13$^{+168.57}_{-102.74}$ & 145.63$^{52.03}_{-31.71}$ & 28.13$^{+1.47}_{-1.47}$ & 5.14$^{+0.25}_{-0.26}$ & 0.84$\pm$0.08 & 1.79\\
HB4 & 18.20$\pm$0.35 & 15.14$\pm$1.09 & 1112.75$^{+238.95}_{-16.86}$ & 317.63$^{+73.75}_{+51.50}$ & 32.70$^{+1.64}_{-1.64}$ & 4.61$^{+0.21}_{-0.21}$ & 0.68$\pm$0.06 &  1.87\\
NB1 & 18.62$\pm$0.41 & 20.42$\pm$0.88 & 1468.04$^{+219.56}_{-173.55}$ & 409.10$^{+67.71}_{-53.54}$ & 33.83$^{+1.10}_{-1.11}$ & 4.44$^{+0.23}_{-0.24}$ & 0.61$\pm$0.06 & 1.89\\
NB2 & 16.60$\pm$0.39 & 18.20$\pm$1.19 & 1626.58$^{+177.68}_{-113.10}$ & $461.28^{+54.34}_{-34.91}$ & 36.62$^{+2.06}_{-2.05}$ & 3.72$^{+0.32}_{-0.28}$ & 0.47$\pm$0.08 & 1.88\\
NB3& 14.79$\pm$0.33 & 14.13$\pm$0.63 & 1049.09$^{+175.96}_{-131.70}$ & 306.09$^{+54.31}_{-40.65}$ & 32.84$^{+1.22}_{-1.22}$ & 4.08$^{+0.32}_{-0.30}$ & 0.48$\pm$0.07  &1.85\\
NB4 & 13.49$\pm$0.30 & 12.30$\pm$0.42 & 916.23$^{+158.92}_{-118.61}$ & 272.13$^{+49.03}_{-36.61}$ & 33.64$^{+1.31}_{-1.30}$ & 3.87$^{+0.37}_{-0.37}$ & 0.41$\pm$0.06 & 1.83\\
FB1 & 12.59$\pm$0.28 & 13.80$\pm$0.54 & 1110.26$^{+276.82}_{-185.03}$ & 328.57$^{+85.54}_{-57.11}$ & 31.16$^{+2.13}_{-2.11}$ & 3.61$^{+0.67}_{-0.60}$  & 0.34$\pm$0.12 & 1.84\\
FB2 & 10.25$\pm$0.21 & 8.91$\pm$0.63 & 749.26$^{+212.02}_{-138.31}$ & 233.34$^{+65.44}_{-42.69}$ & 24.02$^{+2.26}_{-2.46}$ & 3.40$^{+1.10}_{-0.79}$ & 0.30$\pm$0.16 & 1.79\\
       \hline
    \end{tabular}
   
    \label{tab:flux}
\end{table*}

\begin{table*}[ht]
 \caption{Fit-Statistics to compare 3 different models used to describe the X-ray spectra of the source. We have compared Model-3 with Model-1 and Model-2 by computing F-test chance improvement probabilities}.
    \begin{tabular}{llllll}
    \hline
   Segment & $\chi^2/dof (Model-4)$   & $\chi^2/dof$ (Model-1)   & $\chi^2/dof$ (Model-2) &  F-test Prob (Model-4 vs Model-1) & F-test (Model-4 vs Model-2)\\
   \hline
   \hline
HB1 & 661/632 & 686/633 & 696/634 & $1.3\times 10^{-6}$ & $8.30 \times 10^{-8}$ \\
HB2 & 664/632 & 697/633 & 690/634 & $3.1\times 10^{-8}$ & $5.3 \times 10^{-7}$ \\
HB3 & 685/632 & 720/633 & 710/635 & $2.0\times10^{-8}$   & $4.7 \times 10^{-5}$\\
HB4 & 676/632 & 754/634 & 713/635 & $1.1 \times 10^{-16}$ & $2.2\times 10^{-7}$\\
NB1 & 717/632 & 930/634 & 840/635 & $2.0 \times 10^{-36}$ & $1.4 \times 10^{-21}$\\
NB2 & 735/632 & 810/634 & 771/635 & $4.6  \times 10^{-14}$ & $1.2\times 10^{-6}$\\
NB3 & 704/632 & 821/634 & 754/635 & $7.9 \times 10^{-22}$ & $2.0 \times 10^{-9}$\\
NB4 & 712/632 & 814/634 & 761/635 & $4.2 \times 10^{-19}$ & $3.8\times 10^{-9}$\\
FB1 & 658/632 & 725/634 & 684/635 & $4.9 \times 10^{-14}$ & $ 1.9\times 10^{-5}$\\
FB2 & 808/632 & 837/634 & 830/635 & $1.4 \times 10^{-5}$  & $ 7.0\times 10^{-4}$\\

    \hline
    \end{tabular}
   
    \label{tab:stat}
\end{table*}

\begin{table*}[ht]
 \caption{Parameter values obtained by fitting the PDS of  different segments of the
    Z-track.   Lorentzian components and power-law are required to describe the BLN, narrow QPOs and VLFN in the PDS. The parameters of the fit are, power-law index $\alpha$ , break frequency $\nu_{break}$, the characteristics frequency of the  QPOs $\nu_{char}$ ($\nu_{HBO}$ for HBO and $\nu_{NBO}$ for NBO) and  full-width half-maxima $\Delta \nu $.}
    \begin{tabular}{lllllllll}
    \hline
   Segment & $\alpha$ &  $\nu_{break}$  & $\nu_{NBO}$ (significance)   &  $\Delta \nu_{NBO}$ & $\nu_{HBO}$ (significance) & $\Delta \nu_{HBO}$ & $\chi^2/dof$\\
   \hline
   HB1 &  0.74$^{+0.02}_{-0.02}$ & 6.46$^{+0.26}_{-0.24}$  & $-$  &  $-$   &  34.31$^{+0.25}_{-0.25}$ (12.7$\sigma$)  &7.42$^{+0.85}_{-0.76}$ & 122/163\\
   HB2 & 0.85$^{+0.03}_{-0.02}$ & 7.01$^{+0.21}_{-0.20}$  & $-$      &  $-$& 39.06$^{+0.47}_{-0.45}$ (10.8$\sigma$) &   12.33$^{+1.59}_{-1.39}$ & 159/163\\
   HB3 & 0.82$^{+0.02}_{-0.02}$ & 7.24$^{+0.31}_{-0.32}$  & $-$  & $-$    & 41.49$^{+0.71}_{-0.68}$ (5.8 $\sigma$) &   8.45$^{+2.10}_{-1.76}$ & 118/163\\
   HB4& 0.93$^{+0.07}_{-0.06}$& 10.98$^{+1.01}_{-0.90}$ & $-$     & $-$ &    59.51$^{+3.86}_{-3.95}$ (2.6$\sigma$)&   19.95$^{+10.97}_{-9.23}$ & 97/163\\
   NB1 &  1.14$^{+0.09}_{-0.08}$ & $-$                     & 9.62$^{+1.20}_{-0.85}$\footnote{The frequency of LFN}&   11.18$^{+4.15}_{-2.94}$\footnote{FWHM of LFN} &    58.20$^{+5.70}_{-4.92}$\footnote{Frequency of HFN} &   41.11$^{+22.52}_{-14.14}$\footnote{FWHM of HFN} & 143/162 \\
   NB2 &  1.21$^{+0.17}_{-0.17}$ & $-$                    & 6.67$^{+0.16}_{-0.10}$ (3.0$\sigma$) &   2.74$^{+1.65}_{-1.04}$   &    $-$                     &  $-$  &121/165  \\
   NB3 &  1.67$^{+0.11}_{-0.10}$   & $-$                    &6.68$^{+0.06}_{-0.04} $ (5.8$\sigma$)  &   2.45$^{+0.71}_{-0.45}$   &                $-$      & $-$ & 115/165\\
   NB4&  1.36$^{+0.10}_{-0.09}$  & $-$                    & $-$                    & $-$            & $-$  & $-$ & 143/168\\
    FB1 &   1.27$^{+0.15}_{-0.13}$  & $-$ & $-$ & $-$ & $-$ & $-$ & 110/168\\  
    FB2 & 1.69$^{+0.15}_{-0.13}$   & $-$ & $-$ & $-$ & $-$&$-$ & 118/168\\

    \hline
    \end{tabular}
   
    \label{tab:pds}
\end{table*}
\begin{table*}[ht]
 \caption{The table provides  the rms strength (in \%) of different power spectral features (VLFN, BLN and narrow QPOs)  as a function of HID position.}
    \begin{tabular}{llllllll}
    \hline
  Segments &      VLFN-rms   &       Break-rms           & HBO-rms &                    NBO-rms \\
        \hline
HB1 & 6.86$^{+0.32}_{-0.33}$   &   8.49$^{+0.14}_{-0.13}$      & 4.49$^{+0.16}_{-0.18}$ &    $-$\\ 
HB2 & 5.6$^{+0.27}_{-0.23}$    &   8.19$^{+0.12}_{-0.13}$      & 4.04$^{+0.19}_{-0.17}$  &    $-$\\
HB3 &6.57$^{+0.33}_{-0.28}$   &    7.68$^{+0.15}_{-0.15}$      & 2.77$^{+0.23}_{-0.23}$  &    $-$\\
HB4 & 4.25$^{+0.38}_{-0.37}$ &     5.23$^{+0.25}_{-0.24}$      & 2.25$^{+0.42}_{-0.36}$   &    $-$\\
NB1 &2.71$^{+0.37}_{-0.35}$ &               $-$   &   2.82$^{+0.37}_{-0.32}\footnote{HFN-rms}$  &   2.32$^{+0.32}_{-0.25}$\footnote{LFN-rms}\\
NB2 & 2.30$^{+0.27}_{-0.27}$&   $-$       &   $-$             $-$  &                 1.64$^{+0.26}_{-0.25}$\\
NB3 &  1.84$^{+0.17}_{-0.16}$&  $-$       &      $-$       $-$  &                  1.66$^{+0.15}_{-0.14}$\\
NB4 &  1.76$^{+0.12}_{-0.13}$& $-$    & $-$ & $-$ \\
FB1 &  2.41$^{+0.22}_{-0.23}$& $-$ & $-$ & $-$\\
FB2 & 3.35$^{0.40}_{-0.37}$  & $-$ &$-$ & $-$\\
    \hline
    \end{tabular}
   
    \label{tab:rms}
\end{table*}
\begin{table*}[ht]
\caption{Energy dependent rms values (in \%) of BLN and HBO components in the PDS.}
 \begin{tabular}{lllllll}
 Energy-band & \multicolumn{3}{c}{HBO-rms  } & \multicolumn{3}{c}{Break-rms }\\ 
 \hline
 $-$ & HB1 & HB2 & HB3 & HB1 & HB2 & HB3 \\
 \hline
$3-5$ keV &  2.66$^{+0.48}_{-0.39}$ & 2.72$^{+0.66}_{-0.42}$ & $<$ 1.97  & 6.79$^{+0.33}_{-0.46}$ &5.96$^{+0.26}_{-0.30}$ &5.54$^{+0.45}_{-0.54}$\\
$5-8$ keV  &5.59$^{+0.54}_{-0.49}$ & 3.28$^{+0.67}_{-0.54}$ &2.68$^{+0.55}_{-0.37}$ & 10.09$^{+0.42}_{-0.42}$i & 9.38$^{+0.36}_{-0.34}$ &  8.75$^{+0.40}_{-0.40}$\\
$8-15$ keV& 7.73$^{+0.64}_{-0.60}$ & 7.04$^{+0.80}_{-0.78}$& 4.94$^{+1.31}_{-0.92}$ & 14.66$^{+0.37}_{-0.37}$ &13.71$^{+0.43}_{-0.43}$ & 12.20$^{+0.53}_{-0.49}$\\
\hline
    \end{tabular}
    \label{tab:erms}
    
 \end{table*}

\begin{acknowledgement}
Author thanks GH, SAG; DD, PDMSA, and Director URSC for encouragement and continuous support to carry out this research. Author thanks anonymous reviewer for useful and constructive suggestions which helped to improve the quality of the manuscript. 
This work has used the data from the LAXPC Instruments developed at TIFR, Mumbai
and the LAXPC POC at TIFR is thanked for verifying and releasing the data via the ISSDC data archive. We thank the AstroSat Science Support Cell hosted by IUCAA and TIFR for providing the LaxpcSoft software which we used for LAXPC data analysis. This work has used the data from the Soft X-ray Telescope (SXT) developed at TIFR, Mumbai, and the SXT POC at TIFR is thanked for verifying and releasing the data and providing the necessary software tools.
\end{acknowledgement}




\paragraph{Data Availability Statement}
Data underlying this article are available at AstroSat-ISSDC website
\url{https://www.astrobrowse.issdc.gov.in/astro archive/archive}.


\printbibliography
\end{document}